\documentclass[conference]{IEEEtran}

\usepackage{amssymb,amsmath}

\usepackage{gensymb}
\usepackage{xfrac}  
\usepackage{empheq}  
\usepackage[hidelinks,bookmarks=false]{hyperref}
    \hypersetup{colorlinks=false}
\usepackage{cleveref}  
\usepackage{lipsum}  
\usepackage[export]{adjustbox}  
\usepackage{xcolor}  
    \definecolor{highlight}{rgb}{1,1,0}
\usepackage{graphicx}  
\usepackage{grffile}  
\usepackage{tabularx}  
    \newcolumntype{Y}{>{\centering\arraybackslash}X}
\usepackage{multirow}  
\usepackage{colortbl}  
\usepackage{xspace}  
\usepackage{setspace}  
\usepackage{algorithm}  
\usepackage{longtable,tabu}  
\usepackage[noend]{algpseudocode}  
    \algrenewcommand\alglinenumber[1]{{\sffamily\color[rgb]{0.15, 0.35, 0.9}\footnotesize#1}}  
    \algrenewcommand{\algorithmiccomment}[1]{\hfill #1}  
    
    \algblock{ParFor}{EndParFor}
    \algnewcommand\algorithmicparfor{\textbf{parfor}}
    \algnewcommand\algorithmicpardo{\textbf{do}}
    \algnewcommand\algorithmicendparfor{\textbf{end\ parfor}}
    \algrenewtext{ParFor}[1]{\algorithmicparfor\ #1\ \algorithmicpardo}
    \algtext*{EndParFor}{} 
    
    \algblock{Streaming}{EndStreaming}
    \algnewcommand\algorithmicstreaming{\textbf{streaming}}
    \algnewcommand\algorithmicstreamingdo{\textbf{}}
    \algnewcommand\algorithmicendstreaming{\textbf{end\ streaming}}
    \algrenewtext{Streaming}[1]{\algorithmicstreaming\ #1\ \algorithmicstreamingdo}
    \algtext*{EndStreaming}{} 
\usepackage{xpatch}  
    \makeatletter
    \xpatchcmd{\algorithmic}{\itemsep\z@}{\itemsep=2 pt}{}{}
    \makeatother
\usepackage{tikz}
    
    \usetikzlibrary{shapes.misc,positioning,fit,calc}
    
\usepackage[version=4]{mhchem}  
\usepackage{cuted}  
\usepackage{booktabs}  
\usepackage{multirow}  
\usepackage{makecell}  
    
\usepackage{enumitem}  
    \setlist{nosep}  
\usepackage{placeins} 
\usepackage{adjustbox}  
\usepackage{soul}  
\usepackage{etoolbox}  
    \def\Xchannel{arxiv}
    
    \newcommand{\channelifthenelse}[3]{
        \ifdefstring{\Xchannel}{#1}{#2}{#3}
    }
    
    \newcommand{\channelifnotthen}[2]{
        \ifdefstring{\Xchannel}{#1}{}{#2}
    }

\newtheorem{definition}{Definition}
    \AfterEndEnvironment{definition}{\noindent\ignorespaces}

\makeatletter
    \newcommand{\smallotimes}{\mathbin{\mathpalette\make@small\otimes}}
    \newcommand{\make@small}[2]{%
    \vcenter{\hbox{%
        \scalebox{0.6}{$\m@th#1#2$}%
    }}%
    }
    \let\sv@thm\@thm
    \def\@thm{\vspace{0.75em}\let\indent\relax\sv@thm}
\makeatother

\newcommand{\subparagraph}{}
\usepackage{titlesec}
    \titlespacing*{\section}{0pt}{1.0ex}{1.0ex}
    \titlespacing*{\subsection}{0pt}{.75ex}{0.75ex}

\newcommand{\WISHLIST}[1]{}
\newcommand{\REM}[1]{}

\newcommand{\inspaceof}[2]{\phantom{#1}\mathllap{#2}}
\newcommand{\onthefly}{on-the-fly\xspace}
\newcommand{\x}[1]{\mathbf{#1}_\times}

\newcommand{\badge}[1]{
    \trimbox{0mm 0.8mm 2mm 0mm} { 
        \begin{tikzpicture}
            \node (1) [draw, rounded rectangle] {{\tiny\sffamily\bfseries #1}};
        \end{tikzpicture}
    }
}

\newcommand{\warpSize}{\textsc{warpSize}}
\newcommand{\lane}{\textsc{lane}}

\begin{document}

\title{A High-Throughput Solver for Marginalized Graph Kernels on GPU}

\author{
  \IEEEauthorblockN{Yu-Hang Tang, Oguz Selvitopi, Doru Thom Popovici, Ayd\i n Bulu\c{c}}
  \IEEEauthorblockA{
    Computational Research Division, Lawrence Berkeley National Laboratory\\
    Email: \{\href{mailto:tang@lbl.gov}{tang}, roselvitopi, dtpopovici, abuluc\}@lbl.gov
  }
}

\maketitle

\begin{abstract}

We present the design and optimization of a linear solver on General Purpose GPUs for the efficient and high-throughput evaluation of the marginalized graph kernel between pairs of labeled graphs.
The solver implements a preconditioned conjugate gradient (PCG) method to compute the solution to a generalized Laplacian equation associated with the tensor product of two graphs.
To cope with the gap between the instruction throughput and the memory bandwidth of current generation GPUs, our solver forms the tensor product linear system on-the-fly without storing it in memory when performing matrix-vector dot product operations in PCG.
Such on-the-fly computation is accomplished by using threads in a warp to cooperatively stream the adjacency and edge label matrices of individual graphs by small square matrix blocks called tiles, which are then staged in registers and the shared memory for later reuse.
Warps across a thread block can further share tiles via the shared memory to increase data reuse.
We exploit the sparsity of the graphs hierarchically by storing only non-empty tiles using a coordinate format and nonzero elements within each tile using bitmaps.
Besides, we propose a new partition-based reordering algorithm for aggregating nonzero elements of the graphs into fewer but denser tiles to improve the efficiency of the sparse format.

We carry out extensive theoretical analyses on the graph tensor product primitives for tiles of various density and evaluate their performance on synthetic and real-world datasets.
Our solver delivers three to four orders of magnitude speedup over existing CPU-based solvers such as GraKeL and GraphKernels.
The capability of the solver enables kernel-based learning tasks at unprecedented scales.

\end{abstract}

\maketitle

\section{Introduction}\label{introduction}

\begin{figure}[htp!]
  \centering
  \caption{Image and voice recordings are \emph{discretizable} objects in the sense that numeric representations for them can be acquired by sampling at a certain resolution, which is a tunable parameter.
  In contrast, molecules and social networks are \emph{non-sequential and discrete} objects, and thus are better represented by graphs.
  \label{fig:discretizable-vs-discrete}}
  \vspace{0.5em}
  \includegraphics[width=\columnwidth]{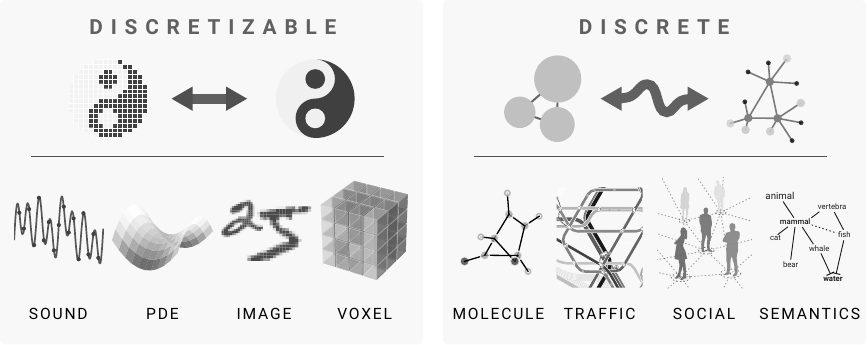}
\end{figure}

Recent advances in machine learning have sparked unique opportunities for building artificial intelligence on graphs, which is a versatile data structure for representing non-sequential data of discrete nature.
As illustrated by \Cref{fig:discretizable-vs-discrete}, a distinction of graph-based discrete data from vector-based discretizable data is that the former consists of indivisible elements that must be inserted or withdrawn atomically. In contrast, the latter consist of discretized samples drawn from a continuous signal at tunable resolutions.
Consequently, graph data does not trivially permit interpolation, convolution, and inner product, which are the operations commonly used in feature extraction.
As a result, special care must be taken to generalize machine learning algorithms that operate on fixed-length feature vectors and uniform grids to their graph-based counterparts.

One way to interface graph data to machine learning algorithms is to apply the \emph{kernel trick}.
A \emph{graph kernel} in this context refers to a function that performs inner product operations between graphs after implicitly transforming them into high- and even infinite-dimensional feature vectors.
A valid graph kernel must be positive definite, meaning that the feature space must be a reproducing kernel Hilbert space.
The inner product thus naturally induces a measure of graph similarity using the cosine of angles in the feature space.
Graph kernels allow a wide range of kernel-based learning methods, \textit{e.g.} support vector machine, Gaussian process regression, spectral clustering, principal component analysis, to operate straightforwardly on graph-based datasets.

\begin{figure*}[!hpt]
  \centering
  \caption{Unlike kernels which compute the inner product between fixed-length explicit feature vectors, the marginalized graph kernel computes the inner product between labeled graphs. Such inner product is defined as the expectation of the inner products between all the simultaneous random walk paths on a pair of graphs, and can be efficiently computed by solving a linear system associated with the tensor product of the two graphs.\label{fig:graph-kernel-visualization}}
  \vspace{0.5em}
  \includegraphics[width=\textwidth]{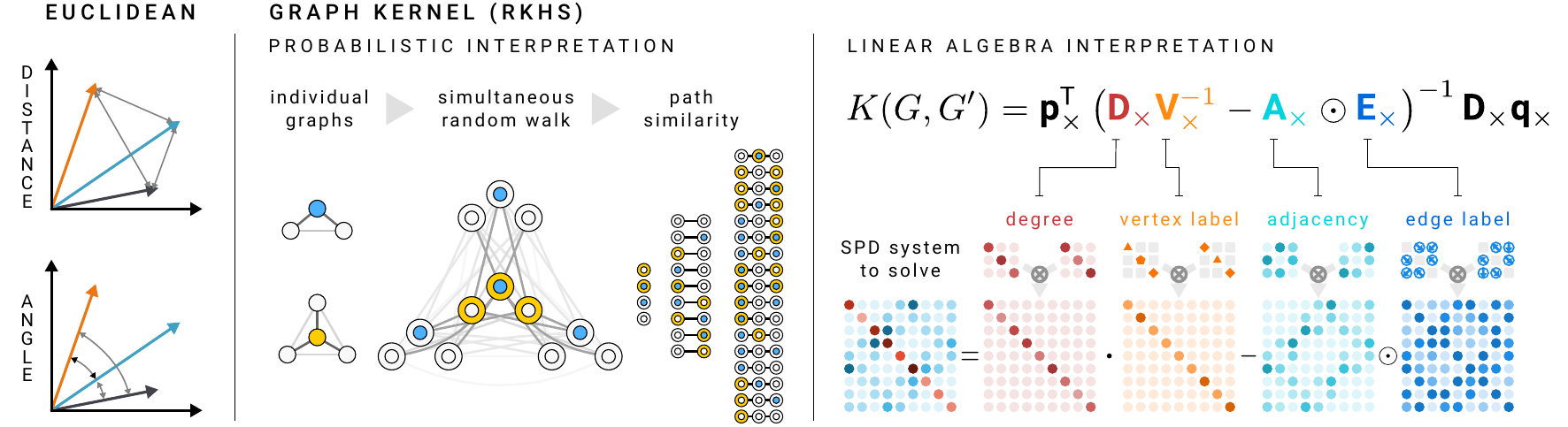}
\end{figure*}

The marginalized graph kernel~\cite{kashimaMarginalizedKernelsLabeled2003} is a powerful tool for graph similarity comparison between labeled and weighted graphs of arbitrary size and topology.
As illustrated in \Cref{fig:graph-kernel-visualization}, the kernel constructs a feature space containing infinitely many dimensions, each of which represents a path on a graph.
The weight of a feature is set equal to the probability of its path in a Markovian random walk process induced by the graph's adjacency matrix.
The overall similarity is then defined as the expectation of partial similarities between all pairs of same-length paths,
each of which is computed as the product of a sequence of node-by-node and edge-by-edge comparisons.
Besides computing an overall similarity score between two graphs, the kernel also defines a measure of node-wise similarity,
which is the expectation of the similarities between all pairs of paths originating from a given pair of nodes.
The node-wise similarity is particularly useful for learning tasks involving the transfer of node labels.
The kernel has found successful application in tasks such as prediction of molecular energy \cite{tangPredictionAtomizationEnergy2019} and protein function \cite{borgwardtProteinFunctionPrediction2005a}.

This paper focuses on the efficient and high-throughput computation of the marginalized graph kernel, which is critical for computing
the pairwise similarity matrix between \emph{all pairs} of graphs, a task that occurs \emph{repeatedly} when training many kernel-based methods.
As will be shown later in \Cref{section:marginalized-graph-kernel} and \channelifthenelse{ipdps}{the online Appendix~\cite{tangHighThroughputSolverMarginalized2019}}{\Cref{section:derivation-of-the-linear-algebra-form-of-the-marginalized-graph-kernel}}, each marginalized graph kernel evaluation between a pair of graphs involves solving a linear system whose size is the product of the number of nodes of the two graphs.
To obtain a pairwise similarity matrix for a dataset of 2000 graphs, each with 100 nodes, we need to solve a million $10^4 \times 10^4$ linear systems.
Thus, a high-performance and high-throughput solver is crucial for applying and scaling the marginalized graph kernel to large datasets.

In this paper, we present a series of algorithms and optimizations, such as \onthefly Kronecker product matrix-vector multiplication, partition-based graph reordering, and sparsity exploitation, to accelerate the marginalized graph kernel computation.
The synergy of the algorithms leads to a solver that achieves a significant performance boost over existing packages on general-purpose graphics processing units (GPGPUs).

The rest of the paper is organized as follows.
In \Cref{section:theoretical-background}, we briefly review related mathematical background knowledge, introduce the formulation of the marginalized graph kernel, and carry out a preliminary analysis to identify design challenges.
In \Cref{section:on-the-fly-XMV}, we explore several design options of a dense Kronecker product matrix-vector multiplication primitive and identify the optimal one through Roofline analyses and microbenchmarking.
In \Cref{section:explicit-sparsity-exploitation}, we examine data structure designs, graph reordering algorithms, and sparse Kronecker product matrix-vector multiplication primitives in order to exploit the sparsity in the graph.
In \Cref{section:tile-sharing-and-load-balancing}, we present data sharing and load balancing approaches for scaling the algorithm onto entire GPUs.
Benchmark datasets and results are given in \cref{section:dataset} and \cref{section:benchmark-result}, respectively.
We discuss the connections between our project and previous ones in \cref{section:related} and conclude the paper in \cref{section:conclusion}.

\section{Theoretical Background}
\label{section:theoretical-background}

\subsection{Preliminaries and Notations}

We use lower case letters in bold font, \textit{e.g.} $\mathbf{a}$, to denote vectors, and upper case letters in bold font, \textit{e.g.} $\mathbf{A}$, to denote matrices. By default, we assume vectors are column vectors. We use $\mathbf{diag}(\mathbf{a})$ to denote a diagonal matrix whose diagonal elements are specified by $\mathbf{a}$. We use \emph{vertex} and \emph{node} interchangeably to refer to the fundamental units of graphs.

\begin{definition}{Undirected graph}\\
An undirected graph $G$ is a discrete structure consisting of a set of uniquely-indexed vertices $V = \{ v_1, v_2, \ldots, v_n \}$ and a set of undirected edges $E \subset V \times V$. The vertices and edges may be labeled using elements from label sets $\Sigma_\mathrm{v}$ and $\Sigma_\mathrm{e}$, respectively.
\end{definition}

\begin{definition}{Weighted graph}\\
In a weighted graph, each edge $(v_i, v_j)$ is associated with a non-negative weight $w_{ij}$. In undirected graphs $w_{ij} = w_{ji}$. $w_{ij} = 0$ if $v_i$ and $v_j$ are not connected by an edge. An unweighted graph can be regarded as a specialized weighted graph where $w_{ij} = 1$ between each pair of $(v_i, v_j)$ connected by an edge and $0$ elsewhere.
\end{definition}

\begin{definition}{Walk on graph}\\
Two vertices are neighbors if they are connected by an edge. A walk on a graph is a sequence of vertices and edges such that all consecutive pairs of vertices are neighbors.
\end{definition}

\begin{definition}{Adjacency matrix}\\
\label{def:adjacency-matrix}
The adjacency matrix of a graph of $n$ vertices is a matrix $\mathbf{A} \in \mathbb{R}^{n \times n}$ with $\mathbf{A}_{ij} = w_{ij}$. The adjacency matrices of undirected graphs are symmetric since $w_{ij} \equiv w_{ji}$.
\end{definition}

\begin{definition}{Edge label matrix}\\
\label{def:edge-label-matrix}
The edge label matrix of a graph of $n$ vertices is a matrix $\mathbf{E} \in {\Sigma_\mathrm{e}}^{n \times n}$ with $\mathbf{E}_{ij} = e_{ij}$. $\mathbf{E}$ has the same symmetry and sparsity pattern with $\mathbf{A}$.
\end{definition}

\begin{definition}{Kronecker product}\label{def:kronecker}\\
Given matrices $\mathbf{A} \in \mathbb{R}^{n \times m}$ and $\mathbf{B} \in \mathbb{R}^{n' \times m'}$, the Kronecker product $\mathbf{P} = \mathbf{A} \otimes \mathbf{B} \in \mathbb{R}^{n n' \times m m'}$ is defined as:
\begin{align*}
  \mathbf{P} = \mathbf{A} \otimes \mathbf{B} \coloneqq
  \begin{bmatrix}
    \mathbf{A}_{1,1} \mathbf{B} & \mathbf{A}_{1,2} \mathbf{B} & \ldots & \mathbf{A}_{1,m} \mathbf{B} \\
    \mathbf{A}_{2,1} \mathbf{B} & \mathbf{A}_{2,2} \mathbf{B} & \ldots & \mathbf{A}_{2,m} \mathbf{B} \\
    \vdots & \vdots & \ddots & \vdots      \\
    \mathbf{A}_{n,1} \mathbf{B} & \mathbf{A}_{n,2} \mathbf{B} & \ldots & \mathbf{A}_{n, m} \mathbf{B}
  \end{bmatrix}
\end{align*}

To better visualize the correspondence between an element of the Kronecker product matrix and its source elements from the operand matrices, we use a quadruple index notation $\mathbf{P}_{ii', jj'}$, which is located at the $(i\times n + i')$-th row and $(j \times m + j')$-th column of $\mathbf{P}$, to denote the element formed by $\mathbf{A}_{ij} \cdot \mathbf{B}_{i'j'}$. Similarly, for a vector $\mathbf{p} = \mathbf{a} \otimes \mathbf{b}$, we use $\mathbf{p}_{ii'}$ to denote its $(i \times n + i')$-th component which is formed by $\mathbf{a}_i \cdot \mathbf{b}_{i'}$.
\end{definition}

\begin{definition}{Generalized Kronecker product}\label{def:generalized-kronecker}\\
Given a set $\mathbb{S}$ whose elements are necessarily numeric, a generalized Kronecker product $\mathbf{P} = \mathbf{A} \overset{\scriptscriptstyle \kappa}{\smallotimes} \mathbf{B}$ between two matrices $\mathbf{A} \in \mathbb{S}^{n \times m}$ and $\mathbf{B} \in \mathbb{S}^{n' \times m'}$ with respect to a kernel $\kappa: \mathbb{S} \times \mathbb{S} \rightarrow \mathbb{R}^{+}$, is a real matrix $\mathbf{P} \in \mathbb{R}^{nn' \times mm'}$ where $\mathbf{P}_{ii', jj'} = \kappa(\mathbf{A}_{ij}, \mathbf{B}_{i'j'})$. In other words, $\kappa$ is a generalization of the real number multiplication operation as used in the standard Kronecker product on $\mathbb{S}$.
\end{definition}

\begin{definition}{Hadamard (element-wise) product}\label{def:hadamard}\\
The element-wise product, also known as the Hadamard product, between two matrices of the same size $\mathbf{A}, \mathbf{B} \in \mathbb{R}^{m\times n}$ is another matrix $\mathbf{A} \odot \mathbf{B} \in \mathbb{R}^{m\times n}$ with $(\mathbf{A} \odot \mathbf{B})_{ij} \coloneqq \mathbf{A}_{ij}\,\mathbf{B}_{ij}$.
\end{definition}
  
\subsection{Marginalized Graph Kernel}\label{section:marginalized-graph-kernel}

We have previously shown \cite{tangPredictionAtomizationEnergy2019} that the computation to apply  marginalized graph kernel between two labeled graphs $G$ and $G'$ can be simplified into solving a linear system involving a generalized Laplacian of the tensor product graph $G \otimes G'$:
\begin{equation}
K_\mathrm{MG}(G, G') = \x{p}^\mathsf{T} \left( \x{D} \x{V}^{-1} - \x{A} \odot \x{E} \right)^{-1} \x{D} \x{q}.\label{eq:mlgk-symmetric-linear-form}
\end{equation}
Here
\begin{itemize}[label=,leftmargin=1em]
    \item $\x{p} \coloneqq \mathbf{p} \otimes \mathbf{p}'$ is the starting probability of a Markovian random walk process from each node of the product graph;
    \item $\x{q} \coloneqq \mathbf{q} \otimes \mathbf{q}'$ is the stopping probability of the random walk process on each node of the product graph;
    \item $\x{A} \coloneqq \mathbf{A} \otimes \mathbf{A}'$ is the adjacency matrix of the product graph;
    \item $\x{D} \coloneqq \mathbf{diag}( \mathbf{d} \otimes \mathbf{d}')$ is the degree matrix of the product graph, while $\mathbf{d}_i = \sum_j \mathbf{A}_{ij} + \mathbf{q}_i$ is the degree of node $i$;
    \item $\x{V} \coloneqq \mathbf{diag}\left( \mathbf{v} \overset{\scriptscriptstyle \kappa}{\smallotimes} \mathbf{v}' \right)$ is a diagonal matrix set by the generalized Kronecker product with respect to a vertex base kernel $\kappa_\mathrm{v}: \Sigma_\mathrm{v} \times \Sigma_\mathrm{v} \rightarrow \mathbb{R}^+$, while $\mathbf{v}$ and $\mathbf{v}'$ contains the vertex labels of $G$ and $G'$, respectively;
    \item $\x{E} \coloneqq \mathbf{E} \overset{\scriptscriptstyle \kappa}{\smallotimes} \mathbf{E}'$ is the generalized Kronecker product between edge label matrices $E$ and $E'$ with respect to an edge base kernel $\kappa_\mathrm{e}: \Sigma_\mathrm{e} \times \Sigma_\mathrm{e} \rightarrow \mathbb{R}^+$.
\end{itemize}
Our extended preprint~\cite{tangHighThroughputSolverMarginalized2019} gives a detailed derivation of \Cref{eq:mlgk-symmetric-linear-form}.
Our earlier work~\cite{tangPredictionAtomizationEnergy2019} contains an example of the rules for determining the specific values for $\mathbf{p}$, $\mathbf{q}$, $\mathbf{V}$, $\mathbf{A}$, and $\mathbf{E}$.

$\mathbf{D_\times}$ and $\x{V}$ are complete diagonal matrices, and $\x{A}$ and $\x{E}$ only have non-zero off-diagonal elements.
The linear system in \Cref{eq:mlgk-symmetric-linear-form} is symmetric and positive definite as long as the base kernels $\kappa_\mathrm{v}(\cdot,\cdot)$ and $\kappa_\mathrm{e}(\cdot,\cdot)$ themselves are positive definite with ranges within $(0,1]$ and $[0,1]$, respectively.
The actual arithmetics involved to compute $\kappa_\mathrm{v}$ and $\kappa_\mathrm{e}$ strongly affect the design of efficient matrix-vector multiplication primitives because they determine the computational costs to generate $\x{E}$ and $\x{V}$ as detailed in \Cref{section:on-the-fly-XMV}.

A degenerate case worth noting is when both the graph nodes and edges are unlabeled.
Such unlabeled graphs eliminate the use of base kernels as well as the $\x{V}$ and $\x{E}$ matrices from \Cref{eq:mlgk-symmetric-linear-form}.
Consequently, \Cref{eq:mlgk-symmetric-linear-form} gets simplified into
\begin{equation}
K_\mathrm{RW}(G,G') = \x{p}^\mathsf{T} \left( \x{D} - \x{A}  \right)^{-1} \x{D} \x{q}. \label{eq:rwgk-symmetric-linear-form}
\end{equation}
\Cref{eq:rwgk-symmetric-linear-form} is essentially the random walk graph kernel proposed by Vishwanathan et al.~\cite{vishwanathanGraphKernels2010}. We denote it as the \emph{unlabeled} graph kernel, and use it as one of the two model problems in performance modeling and algorithm design.

\subsection{Preconditioned Conjugate Gradient Method}

A variety of methods such as conjugate gradient (CG), spectral decomposition, fixed-point iteration, and generalized Sylvester equation can be used to solve the linear system in \Cref{eq:mlgk-symmetric-linear-form} \cite{vishwanathanGraphKernels2010,vishwanathanFastComputationGraph2006}.
Among those, spectral decomposition delivers the best performance \emph{if} the edges are unlabeled or labeled with a small set of distinct elements. CG is favorable in many real-world applications where the edges are labeled using larger and more complex attribute sets.
For example, the edges can be labeled by interatomic distances that span some continuous interval of $\mathbb{R}^+$ when the graphs represent 3D structures of molecules \cite{tangPredictionAtomizationEnergy2019}.
In this case, the spectral decomposition method is no longer advantageous due to the need for looping over all pairs of distinct labels.

\Cref{alg:cg-for-mlgk} illustrates the application of CG, together with a diagonal preconditioner, for solving \Cref{eq:mlgk-symmetric-linear-form}.
Being formulated as an exact solver for symmetric and positive definite linear systems that iteratively minimizes a residual vector in successive orthogonal directions, the method has in practice often being used as an iterative solver because the convergence can be achieved quickly due to the orthogonalization of successive search directions.

\newcommand{\mxMATVEC}[0]{\begin{minipage}[t]{5mm}\includegraphics{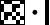}\end{minipage}}
\newcommand{\mxDMATVEC}[0]{\begin{minipage}[t]{5mm}\includegraphics{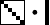}\end{minipage}}
\newcommand{\mxDOT}[0]{\begin{minipage}[t]{3mm}\includegraphics{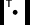}\end{minipage}}
\newcommand{\mxAXPY}[0]{\begin{minipage}[t]{3mm}\includegraphics{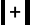}\end{minipage}}

\begin{algorithm}[htp!]
  \caption[Preconditioned conjugate gradient algorithm for the marginalized graph kernel.]{
    Preconditioned conjugate gradient algorithm for the marginalized graph kernel. Legend:
      \mxMATVEC\,: \ off-diagonal symmetric matrix-vector multiplication,
      \mxDMATVEC\,:\ diagonal matrix-vector multiplication,
      \mxDOT\,:    \ vector dot product,
      \mxAXPY\,:   \ (scaled) vector addition.\label{alg:cg-for-mlgk}
  }

  \small
  \begin{algorithmic}[1]
    \Function{CG4GK}{$\mathbf{d}$,$\mathbf{d}'$,$\mathbf{v}$,$\mathbf{v}'$,$\mathbf{A}$,$\mathbf{A}'$,$\mathbf{E}$,$\mathbf{E}'$, $\mathbf{q}$,$\mathbf{q}'$}
    \State $\phantom{x}\mathclap{\mathbf{M}} \ \gets \mathbf{diag}\left[(\inspaceof{\mathbf{A}}{\mathbf{d}} \otimes \inspaceof{\mathbf{A}}{\mathbf{d}}' ) \odot ( \inspaceof{\mathbf{E}}{\mathbf{v}} \overset{\scriptscriptstyle \kappa}{\smallotimes} \inspaceof{\mathbf{E}}{\mathbf{v}}' )^{-1} \right]$ \Comment{\mxAXPY}
    \State $\phantom{x}\mathclap{\mathbf{x}} \ \gets \mathbf{0}$ \Comment{\mxAXPY}
    \State $\phantom{x}\mathclap{\mathbf{r}} \ \gets ( \mathbf{d} \otimes \mathbf{d}' ) \cdot ( \mathbf{q} \otimes \mathbf{q}' )$ \Comment{\mxDMATVEC}
    \State $\phantom{x}\mathclap{\mathbf{z}} \ \gets \mathbf{v} \overset{\scriptscriptstyle \kappa}{\smallotimes} \mathbf{v}'$ \Comment{\mxAXPY}
    \State $\phantom{x}\mathclap{\mathbf{p}} \ \gets \mathbf{z}$ \Comment{\mxAXPY}
    \State $\phantom{x}\mathclap{\rho}       \ \gets \mathbf{r}^\mathsf{T} \mathbf{z}$ \Comment{\mxDOT}
    \Repeat
      \State $\phantom{x}\mathclap{\mathbf{a}} \ \gets ( \inspaceof{\mathbf{A}}{\mathbf{d}} \otimes \inspaceof{\mathbf{A}}{\mathbf{d}}' ) \odot ( \inspaceof{\mathbf{E}}{\mathbf{v}} \overset{\scriptscriptstyle \kappa}{\smallotimes} \inspaceof{\mathbf{E}}{\mathbf{v}}' )^{-1} \cdot \mathbf{p}$ \Comment{\mxDMATVEC}

      \algstore{cg-for-mlgk}
  \end{algorithmic}

  {\setlength{\fboxsep}{0pt}\colorbox{black!10}{\begin{minipage}{\columnwidth}
  
      \begin{algorithmic}[1]
        \algrestore{cg-for-mlgk}
        \State $\phantom{x \ \,\,}$ $ + ( \mathbf{A} \otimes \mathbf{A}' ) \odot ( \mathbf{E} \overset{\scriptscriptstyle \kappa}{\smallotimes} \mathbf{E}' ) \cdot \mathbf{p}$ \label{alg:line:CG-hotspot} \Comment{\mxMATVEC}
        \algstore{cg-for-mlgk}
      \end{algorithmic}
  \end{minipage}}}

  \begin{algorithmic}[1]
      \algrestore{cg-for-mlgk}
      \State $\phantom{x}\mathclap{\alpha}     \ \gets \rho / ( \mathbf{p}^\mathsf{T} \mathbf{a} )$ \Comment{\mxDOT}
      \State $\phantom{x}\mathclap{\mathbf{x}} \ \gets \mathbf{x} + \alpha \mathbf{p}$ \Comment{\mxAXPY}
      \State $\phantom{x}\mathclap{\mathbf{r}} \ \gets \mathbf{r} - \alpha \mathbf{a}$ \Comment{\mxAXPY}
      \State $\phantom{x}\mathclap{\mathbf{z}} \ \gets \mathbf{M}^{-1} \mathbf{r}$ \Comment{\mxAXPY}
      \State $\phantom{x}\mathclap{\rho'}      \ \gets \mathbf{r}^\mathsf{T} \mathbf{z}$ \Comment{\mxDOT}
      \State $\phantom{x}\mathclap{\beta}      \ \gets \rho' / \rho$
      \State $\phantom{x}\mathclap{\mathbf{p}} \ \gets \mathbf{z} + \beta \mathbf{p}$ \Comment{\mxAXPY}
      \State $\phantom{x}\mathclap{\rho}       \ \gets \rho'$
    \Until $\mathbf{r}^\mathsf{T} \mathbf{r} < \epsilon$
    \State \Return $\mathbf{x}$
    \EndFunction
  \end{algorithmic}
\end{algorithm}

\subsection{Preliminary Roofline Analysis}
\label{section:preliminary-roofline-analysis}

\begin{figure}[htb!]
  \centering
  \caption{A Roofline analysis shows that the Kronecker product matrix-vector multiplication operation in the conjugate gradient solver for the marginalized graph kernel is memory-bound if implemented na\"ively on the Volta V100 GPU. A possible solution is to regenerate the product matrix \onthefly without storing it. Hardware metrics provided by \cite{jiaDissectingNVIDIAVolta2018}. Vertical lines correspond to the \onthefly  solver that uses each element $r = 4, 16, 64$ times when computing the matrix-vector product for solving \Cref{eq:rwgk-symmetric-linear-form} where $E=0, F=4, X=3$.\label{fig:roofline-theoretical}}
  \includegraphics[width=\columnwidth]{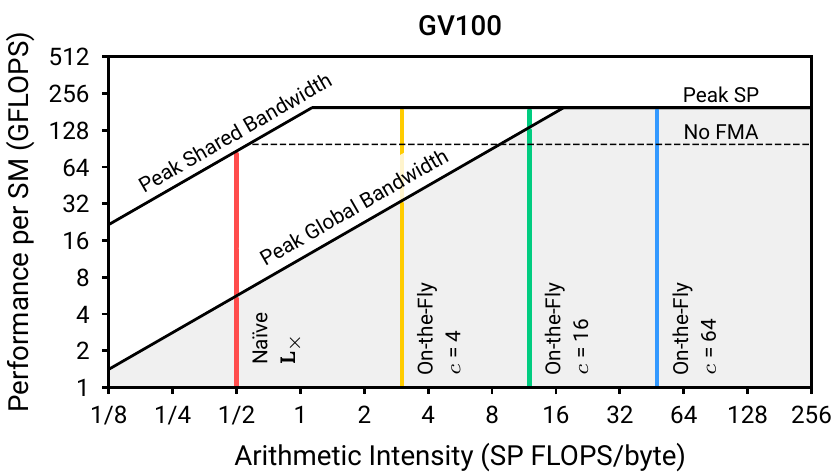}
\end{figure}

The Kronecker product matrix-vector multiplication operation $( \mathbf{A} \otimes \mathbf{A}' ) \odot ( \mathbf{E} \overset{\scriptscriptstyle \kappa}{\smallotimes} \mathbf{E}' ) \cdot \mathbf{p}$, as highlighted on line~\ref{alg:line:CG-hotspot} of \Cref{alg:cg-for-mlgk}, has the highest order of asymptotic complexity of $\mathcal{O}(N^2)$ for a $N \times N$ system, and is the hotspot of the CG algorithm.
Hence, we construct a Roofline model \cite{williamsRooflineInsightfulVisual2009} to estimate the potential profitability of accelerating this operation with GPUs.
We will first focus on fully connected graphs and show later in \Cref{section:explicit-sparsity-exploitation} how sparsity and locality in the graph can be exploited to improve performance further.
Motivated by real-world applications that we encounter as exemplified in the Appendix of our extended preprint~\cite{tangHighThroughputSolverMarginalized2019},
we use an abstract model for the storage and arithmetic cost of the computation. In this model, we assume that a floating-point number occupies $F$ bytes, an edge label occupies $E$ bytes, and a function evaluation of $\kappa_\mathrm{e}(\cdot, \cdot)$ costs $X$ floating-point operations.

In a na\"ive implementation, the product matrix $\x{L} \doteq ( \mathbf{A} \otimes \mathbf{A}' ) \odot ( \mathbf{E} \overset{\scriptscriptstyle \kappa}{\smallotimes} \mathbf{E}' )$ is precomputed beforehand and reused in the CG loop.
Given a pair of graphs each with $n$ and $m$ nodes, respectively, the na\"ive solver needs to load a floating-point matrix $\x{L}$ of $nm\times nm$ elements and a right-hand side vector $\mathbf{p}$ of $nm$ elements, and perform $n^2 m^2$ floating point fused multiply-additions.
Hence, the arithmetic intensity of the na\"ive solver is $2 n^2 m^2 / (n^2 m^2 F + n m F) \rightarrow \frac{2}{F}$, or $\frac{1}{2}$ in single precision mode.
On the Volta GPU architecture, the solver is severely memory-bound, achieving at most $3\%$ utilization of the peak floating-point performance as predicted by the Roofline plot in \Cref{fig:roofline-theoretical}.

A further disadvantage of the na\"ive approach is that the product matrix takes up a prohibitively large amount of storage space.
Such behavior could limit both the size of the graphs as well as the concurrency of pairwise graph kernel computations that a GPU can accommodate.

\begin{algorithm}[htp!]
  \caption{A high-level outline of the \onthefly Kronecker product matrix-vector multiplication (XMV) algorithm.\label{alg:outline-algorithm}}
  \small
  \begin{algorithmic}[1]
    \Function{XMV}{$(\mathbf{A},\mathbf{A}')$, $(\mathbf{E},\mathbf{E}')$, $\mathbf{a}$, $\mathbf{p}$}

      \Streaming{length-$c$ chunks in $\mathbf{A},\mathbf{E}$} \Comment{{\color[rgb]{.5,.5,.5} \scriptsize\sffamily\bfseries OUTER LOOP}}
        \Streaming{length-$c$ chunks in $\mathbf{A}',\mathbf{E}'$} \Comment{{\color[rgb]{.5,.5,.5} \scriptsize\sffamily\bfseries INNER LOOP}}
          
          \For{each $e_{ij}$ in first chunk}
            \For{each $e'_{i'j'}$ in second chunk}
              \State $a_{ii'} \gets a_{ii'} + \kappa(e_{ij}, e'_{i'j'}) \cdot \mathbf{p}_{jj'}$
            \EndFor
          \EndFor\vspace{-2mm}
        \EndStreaming
      \EndStreaming
    \EndFunction
  \end{algorithmic}
  \vspace{-1.5mm}
\end{algorithm}

In \Cref{alg:outline-algorithm}, we outline an \onthefly Kronecker product matrix-vector multiplication (XMV) algorithm that directly computes the inner product $\x{L} \cdot \mathbf{p}$, instead of $\x{L}$, in an attempt to trade data movement with arithmetic operations.
The algorithm takes advantage of the Kronecker product structure of $\x{L}$ to repeatedly recreate the matrix without storing it.
It is achieved by streaming elements from the pair of individual graphs and caching them to perform the computation.
We can perform $c^2$ edge kernel evaluations using only fast memory and registers for every length-$c$ chunk of edge weight/label pairs streamed from the graphics memory.
With a double loop structure, which amortizes the cost of loading one of the two graphs, the \onthefly approach can achieve an arithmetic intensity of $\frac{c^2 X}{c(E + F)} = \frac{c X}{E + F}$, or specifically $\frac{3}{4}c$ in the unlabeled case.
As shown in \Cref{fig:roofline-theoretical}, tuning $c$ can thus be used as a straightforward approach to achieve the highest utilization of the computing power on the Volta GPU.
Note that regenerating the product matrix only increases the constant factor, but does not alter the order of the computational complexity of the matrix-vector multiplication operation.
Therefore, this approach is profitable as long as the gain in instruction throughput outweighs the added cost of base kernel evaluations.

\section{On-the-Fly Kronecker Product Formation and Matrix-Vector Multiplication}\label{section:on-the-fly-XMV}

For dense adjacency and edge label matrices of fully connected graphs, we propose three concrete implementations of the \onthefly XMV algorithm as outlined in \Cref{alg:outline-algorithm} on the Volta GPU. All three primitives adopt a warp-synchronous high-throughput programming model, where every 32 consecutive threads within a warp work cooperatively on a pair of graphs. A na\"ive implementation that uses a precomputed product matrix $\x{L}$ is described in our extended preprint~\cite{tangHighThroughputSolverMarginalized2019}.
We introduce methods to exploit the sparsity of the graphs in \Cref{section:explicit-sparsity-exploitation}, and methods for sharing data and work within a thread block in \Cref{section:tile-sharing-and-load-balancing}.

\begin{figure}[t!]
  \centering
  \caption{In a tile-based \onthefly Kronecker matrix-vector multiplication primitive, the product matrix is regenerated and multiplied with the right-hand side vector by streaming and caching graphs in small pieces called \emph{tiles} in a double loop.\label{fig:kronecker-matvec}}
  \includegraphics[width=\columnwidth]{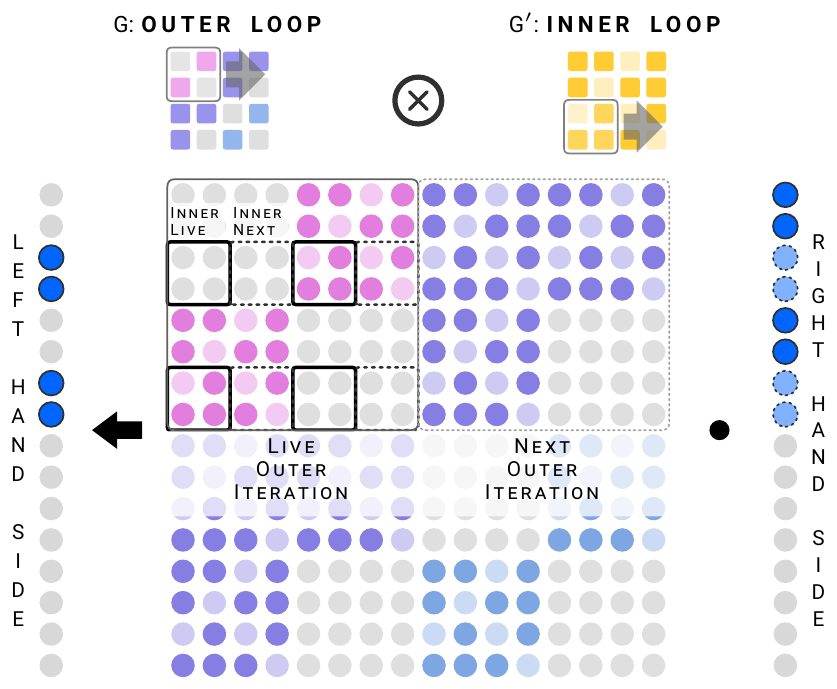}
\end{figure}

We denote the off-chip DDR or HBM memory attached to a GPU as the \emph{device} memory and the on-chip addressable SRAM as the \emph{shared} memory. A \emph{global} load or store operation accesses the device memory, while a \emph{shared} load or store operation accesses the shared memory. 

\begin{figure*}[t!]
  \centering
  \caption{Detailed benchmark and Roofline analysis of the three \onthefly XMV primitives. Each primitive other than the na\"ive one are instantiated with multiple sets of parameters as given underneath each bar. For the \emph{shared tiling} primitive, the two parameters specify the height and width of the tiles that are streamed by the primitive; for the \emph{register blocking} primitive, the two parameters have similar meanings to their shared tiling counterparts; for the \emph{tiling-blocking} primitive, the two parameters corresponds to the size of the square tiles streamed and the length of the chunks staged in registers, respectively.\label{fig:xmv-roofline}}
  \includegraphics[width=\textwidth]{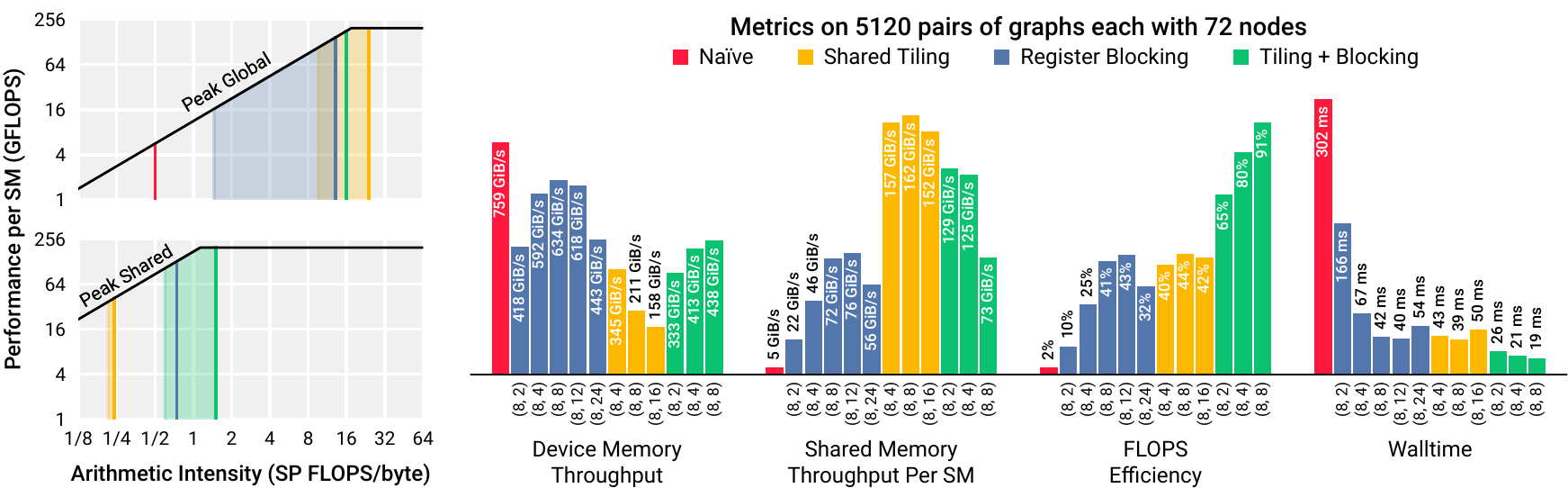}
\end{figure*}

\subsection{Shared Tiling}\label{section:shared-tiling}

Our first implementation, the \emph{shared tiling} primitive, uses the shared memory as a staging area to reduce the usage of global load instructions and device memory bandwidth. \Cref{fig:roofline-theoretical} shows that it is much easier to attain performance close to the theoretical peak on the Volta GPU by loading data from the shared memory, which can provide more than $10^4$ GB/s bandwidth.

As shown in \Cref{fig:kronecker-matvec}, the shared tiling primitive streams $t \times r$ tiles and the corresponding right-hand side elements from the device memory to the shared memory for computation.
The tiles and the right-hand side elements are all loaded cooperatively by a warp to ensure coalesced access.
Among the threads in a warp, the computation for a pair of tiles is parallelized along the rows of the product matrix in a round-robin manner. The work is serialized among the columns within each thread. In the actual code, we choose $t=8$ and explicitly unrolled the row loops by a factor of two to obtain an instruction-level parallelism of two on warps of 32 threads.

\subsection{Register Blocking}\label{section:register-blocking}

Our second implementation, the \emph{register blocking} primitive, uses the register file to stage and reuse matrix elements.
While the work is parallelized along the rows in the same way as the shared tiling primitive, each thread here will independently stream length-$r$ chunks from the rows that it owns to compute $r^2$ elements of the product matrix.
Due to the synchronous execution behavior within CUDA thread warps, threads in a warp can still share right-hand side elements via the shared memory because the march across the columns is lock-stepped between consecutive $t \times r$ blocks.
The primitive is simpler than the shared tiling primitive, but causes a higher register pressure and may generate unnecessary global memory transactions depending on the value of $r$.

\subsection{Combining Shared Tiling and Register Blocking}

Our third implementation combines shared tiling and register blocking. This \emph{tiling-blocking} primitive aims to reduce shared and global memory transactions while simultaneously reducing register pressure.
Here, a $t \times t$ tile is first cached in the shared memory, and then further staged in registers over $r$-element chunks.
It can be implemented easily by placing compiler directives to unroll the inner loops over column indices.

\subsection{Performance Analysis}
\label{section:roofline-analysis}

\newcommand{\BSxLDG}{$n^2 m^2 F$}
\newcommand{\BSxSTG}{$n m F$}
\newcommand{\BSxLDS}{-}
\newcommand{\BSxSTS}{-}
\newcommand{\BSxOPS}{$2 n^2 m^2$}
\newcommand{\BSxSAI}{-}
\newcommand{\BSxGAI}{$\dfrac{2}{F}$}

\newcommand{\RBxLDG}{$n^2 m^2 (\tfrac{t}{r}E+\tfrac{t + r}{r}F)/t^2$}
\newcommand{\RBxSTG}{$n m F$}
\newcommand{\RBxLDS}{$n^2 m^2 F$}
\newcommand{\RBxSTS}{$n^2 m^2 F / t^2$}
\newcommand{\RBxOPS}{$n^2 m^2 X$}
\newcommand{\RBxGAI}{$\dfrac{t^2 X}{\sfrac{t}{r} E + (1 + \sfrac{t}{r})F}$}
\newcommand{\RBxSAI}{$\dfrac{X}{(1 + \sfrac{1}{t^2})F}$}

\newcommand{\STxLDG}{$n^2 m^2 (\tfrac{t}{r}E+\tfrac{r + t}{r}F)/t^2$}
\newcommand{\STxSTG}{$n m F$}
\newcommand{\STxLDS}{$n^2 m^2 (\tfrac{r + 1}{r}E+\tfrac{2r + 1}{r}F)$}
\newcommand{\STxSTS}{$n^2 m^2 (\tfrac{t}{r}E+\tfrac{r + t}{r}F)/t^2$}
\newcommand{\STxOPS}{$n^2 m^2 X$}
\newcommand{\STxGAI}{$\dfrac{t^2 X}{\sfrac{t}{r} E + (1 + \sfrac{t}{r})F}$}
\newcommand{\STxSAI}{$\dfrac{X}{(1 + \sfrac{1}{r}) E + (2 + \sfrac{1}{r}) F}$}

\newcommand{\STRBxLDG}{$n^2 m^2 (E + 2F)/t^2$}
\newcommand{\STRBxSTG}{$n m F$}
\newcommand{\STRBxLDS}{$n^2 m^2 (\tfrac{r + t}{rt}E+\tfrac{r + t}{rt}F)$}
\newcommand{\STRBxSTS}{$n^2 m^2 (E + F)/t^2$}
\newcommand{\STRBxOPS}{$n^2 m^2 X$}
\newcommand{\STRBxGAI}{$\dfrac{t^2 X}{E + 2F}$}
\newcommand{\STRBxSAI}{$\dfrac{X}{(\sfrac{1}{r} + \sfrac{1}{t}) E + (\sfrac{1}{r} + \sfrac{1}{t}) F}$}

\begin{table*}[t!]
  \caption{Operation count, load/store count, and asymptotic arithmetic intensity of the \onthefly Kronecker product matrix-vector multiplication (XMV) operation from line \ref{alg:line:CG-hotspot} of \Cref{alg:cg-for-mlgk} during one CG iteration.
  For a detail derivation of the equations, see \channelifthenelse{ipdps}{the online Appendix~\cite{tangHighThroughputSolverMarginalized2019}}{\Cref{section:primitives}}.
  $n$ and $m$: numbers of nodes of the two graphs, respectively; $E$: byte size of an edge label; $F$:  byte size of an edge weight; $X$: base kernel operation count.\label{table:op-ldst-count}}
  \setlength{\tabcolsep}{6pt}
  
  \begin{tabularx}{1.0\textwidth}{@{}ccYYY@{}}
  \toprule
  &
  \thead[t]{Naive} &
  \thead[t]{$t \times r$ shared tiling} &
  \thead[t]{$t \times r$ register blocking} &
  \thead[t]{length-$r$ register blocking\\within $t \times t$ shared tiling}
  \\
  \midrule

  \thead[c]{Ops.}         & \BSxOPS & \STxOPS & \RBxOPS & \STRBxOPS\\
  \thead[c]{Global Load}  & \BSxLDG & \STxLDG & \RBxLDG & \STRBxLDG \\
  \thead[c]{Global Store} & \BSxSTG & \STxSTG & \RBxSTG & \STRBxSTG\\
  \thead[c]{Shared Load}  & \BSxLDS & \STxLDS & \RBxLDS & \STRBxLDS\\
  \thead[c]{Shared Store} & \BSxSTS & \STxSTS & \RBxSTS & \STRBxSTS\\ \midrule
  \thead[c]{A.I. Global}  &
    {\footnotesize \BSxGAI} &
    {\footnotesize \STxGAI} &
    {\footnotesize \RBxGAI} &
    {\footnotesize \STRBxGAI}
  \\
  \thead[c]{A.I. Shared}   &
    \BSxSAI &
    {\footnotesize \STxSAI} &
    {\footnotesize \RBxSAI} &
    {\footnotesize \STRBxSAI}
  \\
  \bottomrule
  \end{tabularx}
\end{table*}

From \Cref{fig:xmv-roofline}, we can see that the tiling-blocking primitive performs the best in terms of time-to-solution.
It also achieves the best FLOPS efficiency, defined as the ratio between the actual throughput of floating point operations and the theoretical peak after adjusting for FMA percentage.
Hence, the \textbf{tiling-blocking primitive with $t=8$ and $r=8$} is chosen as the building block for subsequent kernels with more optimizations.
We denote the $8 \times 8$ square tiles as \textbf{octiles} hereafter.
The shared tiling primitive and the register blocking primitive performed nearly equally well, yet was not able to achieve the best performance.
The shared tiling primitive is unsurprisingly bound by the shared memory throughput as indicated by the measured shared memory bandwidth utilization and the Roofline model.
The register blocking primitive is bound by global memory throughput when $r$ is small, yet suffers from register spilling right before it reaches the top of the Roofline model with $r = 24$.
Additional tests on a Titan X Pascal graphics card indicate that the shared tiling primitive performs better than the register blocking primitive on accelerator equipped with GDDR memories, but the tiling-blocking primitive still provides the best performance with most balanced utilization of hardware resources.

\WISHLIST{Discussion: which hardware characteristics favor which primitive, max number of graphs on the fly with each primitive (explicit ones are heavily constrained), which label favor which primitive, which edge primitive favor which primitive}

\section{Explicit Sparsity Exploitation}\label{section:explicit-sparsity-exploitation}

Many graphs encountered in real-world applications harbor a certain degree of sparsity, which can be exploited to optimize performance.
For example, a SMILES string represents a molecular graph where edges connect only atoms that are chemically bonded. In this case, the maximum number of edges on each node is capped by the maximum number of bonds that an atom can form, which rarely exceeds 8.
A road network graph is also sparse with 3-way and 4-way junctions dominating the map.
Even for 3D molecular structures where edges encode contact relationships between all pairs of atoms, the graphs can still be sparse due to the spatial locality of non-bond interactions.

We adopt a two-level methodology to exploit the sparsity in the graphs.
In the first level, we exploit the sparsity at the octile granularity by reducing non-empty tiles through graph reordering.
In the second level, we exploit the sparsity within individual octiles by using a compact storage scheme that stores only non-zero elements of the tiles, and by designing corresponding sparse XMV primitives.
In the rest of this section, we use graph and matrix terms interchangeably.

\subsection{Inter-Tile Sparsity}

The sparsity of graphs can be readily exploited within the \onthefly XMV framework by pruning empty tiles that contain no edges.
The implementation of this pruning process as a pre-processing pass is trivial, but its efficiency depends on our ability to find empty tiles in the matrix.
Hence, we resort to reordering algorithms to group the nonzeros of the matrix into as few tiles as possible.
Among a plethora of heuristics in the literature for reordering matrices, the ones that we have experimented with are:
\begin{itemize}

  \item a custom partitioning-based reordering (PBR) algorithm~\cite{Selvitopi2017} that targets explicitly the objective of minimizing the number of non-empty tiles;

  \item the Reverse Cuthill-McKee (RCM) algorithm \cite{georgeComputerSolutionLarge1981a}, which is a heuristic that has found widespread use for fill-in and matrix bandwidth reduction;

  \item a scheme based on solving the Traveling Salesman Problem (TSP) \cite{pinarImprovingPerformanceSparse1999} with heuristics.

  \item a scheme using space-filling curves such as the Morton curve \cite{tangAcceleratingDissipativeParticle2014} or the Hilbert curve when the vertices are known to come from an embedding in a Euclidean space.

\end{itemize}

Among the four reordering methods, we have found that the PBR-based method delivers the most reduction in non-empty octiles using a moderate amount of time.

\begin{figure}[t]
  \centering
  \caption{An example where the partition-based reordering (PBR) method outperforms the amino acid sequence order and the RCM order, yielding graphs with fewer and more densely occupied tiles on two molecular graphs from the PDB dataset.\label{fig:graph-reordering}}
  \includegraphics[width=\columnwidth]{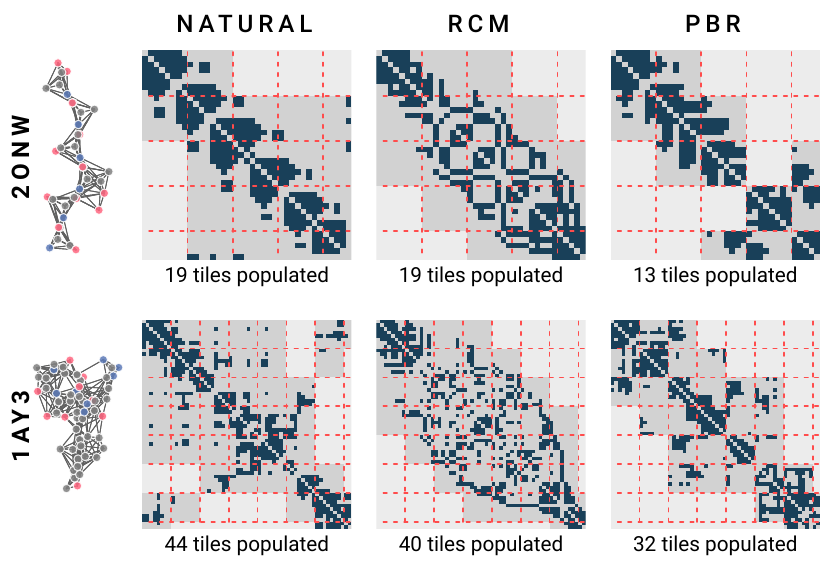}
\end{figure}

The Morton-based method delivers less reduction than RCM and PBR despite being marginally faster.
The TSP-based method achieves a reduction rate between RCM and PBR.
However, the running time of the TSP-based reordering algorithm is substantially longer than all other reordering methods by orders of magnitude.
Hence, we decide to focus only on RCM and PBR in subsequent discussions, and present two examples of molecular graphs representing the protein \texttt{2ONW} and \texttt{1AY3} from the Protein Data Bank (PDB) in their natural orders, the RCM order, and the PBR order, respectively, in Fig.~\ref{fig:graph-reordering}.

\begin{figure}[ht!]
  \centering
  \caption{The PBR method consistently outperforms other graph reordering algorithms and improves the sparsity pattern of graphs from four datasets of real and synthetic graphs. \label{fig:tile-histogram}}
  \includegraphics[width=\columnwidth]{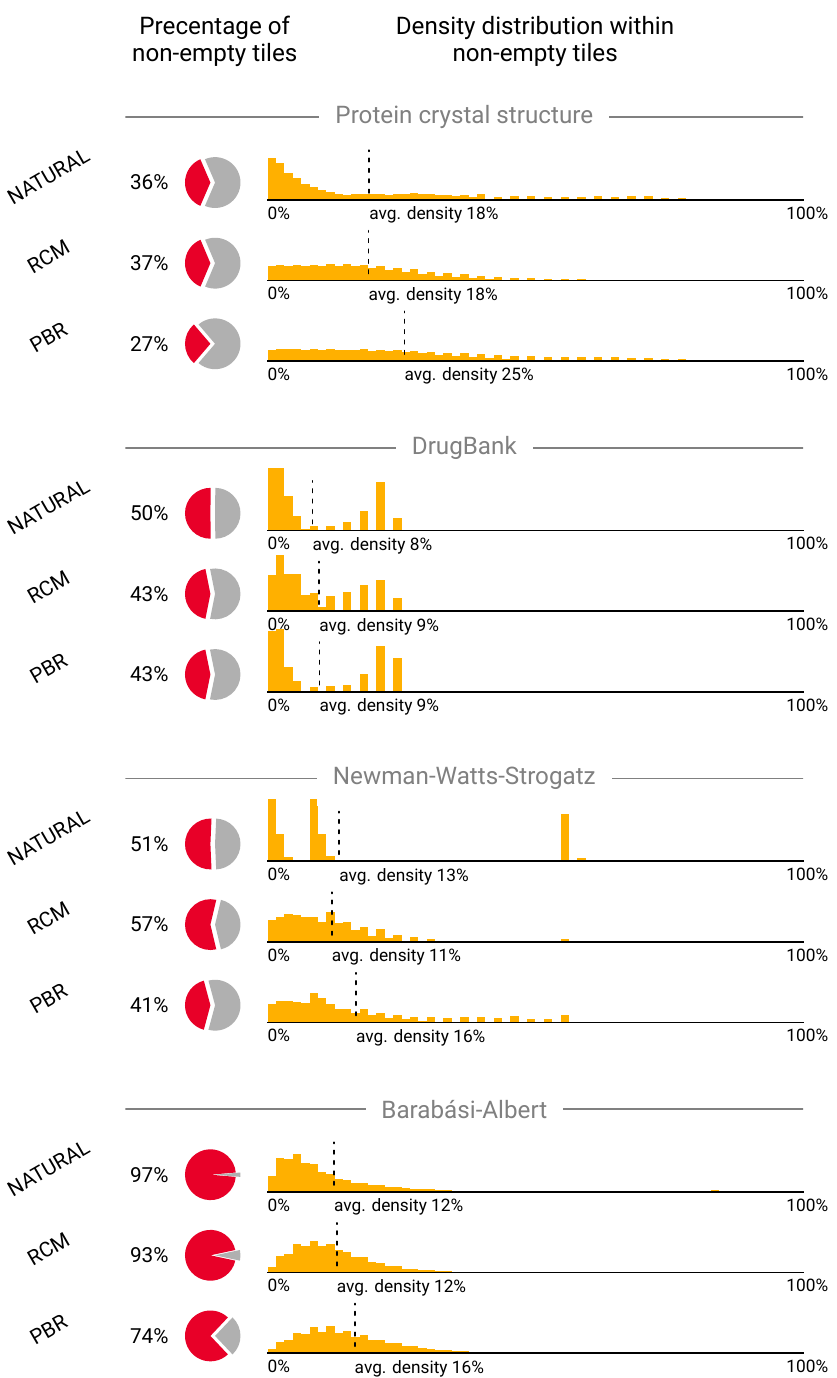}
\end{figure}

In the particular case where the graphs represent 3D protein structures, the nodes in their \emph{natural} order, \textit{e.g.} the order of the corresponding amino acid residues in the primary structure of the protein, already yields nearly optimal sparsity pattern in the adjacency matrix. However, the PBR order can still beat the natural order in reducing non-empty tiles for different datasets, as evident in \Cref{fig:graph-reordering,fig:tile-histogram}.
Moreover, reordering is useful in the general case because the natural orderings of the nodes are not always available.

\noindent
\textbf{Partitioning-based Reordering (PBR) for improved tile density}

The goal of PBR in our case is to reorder nodes in a graph $G=(V, E)$, \textit{i.e.}, to come up with a permutation of the rows and columns of the corresponding matrix, in order to minimize the number of non-empty $t \times t$ square tiles.

Let $\Pi(G)=\{V_1, V_2,\ldots, V_K\}$ be a perfectly balanced $K$-way vertex partition of $G$ with $K=\lceil |V|/t \rceil$, where all parts in $\Pi(G)$ with the possible exception of the last part has exactly the same number of vertices.
$\Pi(G)$ then implies a vertex ordering, where the vertices in $V_k$ are ordered before the vertices in $V_{k+1}$, for $1 \leq k < K$.
Observe that for any $1 \leq k \neq \ell \leq K$, if there is at least one edge between the nodes within $V_k$ and $V_\ell$, then the tile at the intersection of $k$th row stripe and $\ell$th column stripe of the matrix, as well as its symmetric counterpart, are non-empty.
If there are no edges between $V_k$ and $V_\ell$, then the respective tiles are empty.
Therefore, we can define the objective of PBR as finding the $\Pi(G)$ that minimizes
\begin{equation}
  |\{(V_k, V_\ell): k \neq \ell \mbox{ and } (v_i\in V_k, v_j\in V_\ell) \in E \}|.\label{eq:partobj}
\end{equation}

To seek a good $\Pi(G)$, we utilize an approach~\cite{Selvitopi2017} that is fast and has a consistent objective with~\eqref{eq:partobj} but does not always guarantee a perfectly balanced partition.
Perfectly balanced graph partitioning problem has previously been studied in the literature~\cite{sandersThinkLocallyAct2013, Benlic2011}, usually with a different objective of minimizing the number of inter-partition edges.
These approaches also emphasize partition quality over speed and rely on expensive algorithms such as tabu search.
The approach that we use here derives from a recursive bipartitioning scheme that initially aims at reducing the messages sent in a parallel application, which are modeled as off-diagonal blocks in a matrix.
The bipartitioning heuristics are much faster than the algorithms used for perfectly balanced partitioning.

Even though the PBR algorithm \cite{Selvitopi2017} does not guarantee that the partitions be perfectly balanced, an imbalance is rare as long as all vertices have the same weight, which is precisely the case in our work.
Nonetheless, for the cases in which the partitioner could not obtain a perfectly balanced partition, we append an extra refinement step to move vertices from the overloaded part to the underloaded part based on the Fiduccia-Mattheyses (FM) algorithm~\cite{Fiduccia1982}.
We also utilized a custom weight distribution, as opposed to a single imbalance parameter, in the recursive bipartitioning process to promote equally sized parts.
Moreover, we adjust the parameters of the partitioner to ensure a tight
balancing constraint by setting the refinement algorithm to boundary FM with tight balance.
Finally, we set the cost of the message nets, a parameter that emphasizes the importance of
the reduction of non-empty tiles, to a large value such as 50.

\WISHLIST{refer to some table for properties of these datasets}

In Fig.~\ref{fig:tile-histogram}, we illustrate the performance of the PBR order as compared to the natural order and the RCM order on four different datasets as detailed in \Cref{section:dataset}.
PBR achieves the best reduction over the natural ordering in all
datasets, while RCM can only improve the non-empty octile count in two of the datasets.

\noindent
\textbf{Reordering overhead} Reordering is justified when its cost is smaller than the computational savings it enables. The PBR reordering incurs a linear-time pre-processing overhead proportional to the number of non-zeros in the matrices, while the marginalized graph kernel incurs a quadratic cost in the number of non-zeros during each CG iteration.
Moreover, the graph kernel often has to be evaluated on all pairs of graphs for hundreds of times to train a machine learning model, while the training data only need to be reordered once.
Hence, the overhead of the PBR reordering can be quickly amortized and leads to shorter overall time-to-solution.

\subsection{Intra-Tile Sparsity}

As already demonstrated in \Cref{section:on-the-fly-XMV}, the tiling-blocking kernel is very efficient on dense tiles.
Moreover, the kernel is still efficient on most sparse graphs because our reordering algorithms tend to create locally dense areas in the matrices.
Nonetheless, as seen from \Cref{fig:tile-histogram}, although the reordering methods indeed increase the octile density compared to the natural order, the non-empty tiles can still be up to 90\% empty.
Hence, we can attain considerable savings by storing and processing only the nonzero elements instead of treating the tiles as dense.

In order to exploit sparsity within an octile, we use a compact layout to store only nonzero elements.
An accompanying 64-bit integer, whose $i$th element is set if the $i$th element is nonzero, is used to locate the nonzero elements in the original octile.
We then rely on bit manipulations to find the indices of the nonzero elements.
Compared to the dense octile representation, the sparse representation reduces unnecessary global memory transactions besides wasting flops.
However, this comes at the expense of increased shared memory utilization.

\begin{figure}[t]
  \centering
  \caption{The comparative advantages of the different dense/sparse primitives depend on the sparsity patterns of the tiles. Dynamic selection of the primitives can thus lead to performance improvements.\label{fig:crossover}}
  \includegraphics[width=\columnwidth]{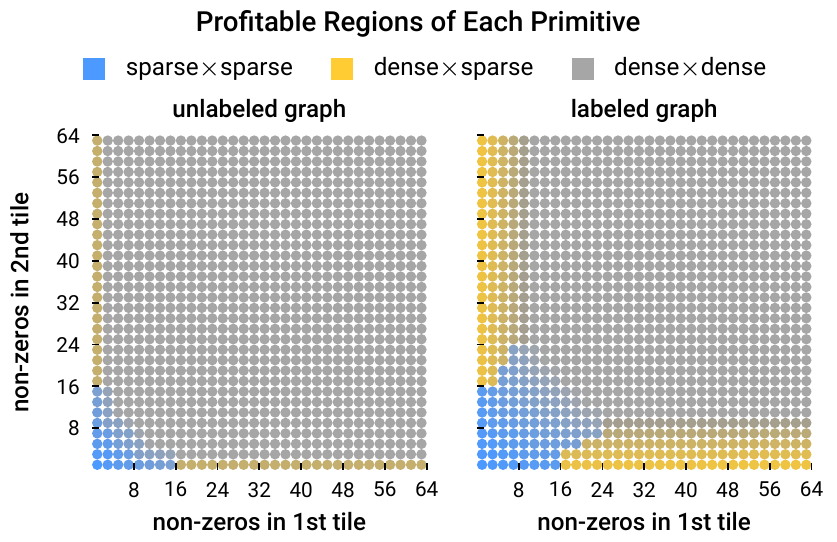}
\end{figure}

\noindent
\textbf{Hybrid dense-sparse computation} The optimal way of evaluating the XMV operation given a pair of tiles depends on the sparsity of the tiles.
Utilizing a single primitive for the entire execution may hurt the performance as the kernels' performance largely depend on the octile density, which can vary significantly within and across datasets as visualized in \Cref{fig:tile-histogram}.
As such, we designed two new types of XMV primitives in addition to the $\mathrm{dense} \times \mathrm{dense}$ kernel:
(i) a primitive for the tensor product between a dense tile and a sparse tile, or vice versa ($\mathrm{dense} \times \mathrm{sparse}$),
and (ii) a primitive for the tensor product between two sparse tiles ($\mathrm{sparse} \times \mathrm{sparse}$).

Fig.~\ref{fig:crossover} illustrates the best performing product kernel for
a varying number of nonzeros of the two source octiles for both labeled and unlabeled graphs.
The $\mathrm{sparse}\!\times\!\mathrm{sparse}$ kernel performs the best when each of the octiles contains up to 8-10 nonzeros for the unlabeled graphs and up to 16 nonzeros for the labeled graphs.
The $\mathrm{dense}\!\times\!\mathrm{dense}$ kernel runs the fastest from that point on as both of the octiles get denser.
In the rest, the $\mathrm{dense}\!\times\!\mathrm{sparse}$ kernel performs better.

\WISHLIST{In order to ease the execution logic, we drop the
  $\mathrm{dense}\!\times\!\mathrm{sparse}$ kernel as it is the worst-performing
  one among the three.}

In our production kernel, we dynamically select either the $\mathrm{sparse}\!\times\!\mathrm{sparse}$ or the $\mathrm{dense}\!\times\!\mathrm{dense}$ kernel before carrying out the tensor product operations depending on the type of the graph and the number of products the two octiles require.
The octiles are always stored in a compact form and expanded in the shared memory after loading them from global memory.

\section{Tile Sharing and Load Balancing}
\label{section:tile-sharing-and-load-balancing}

\subsection{Block-Level Sharing}

To fully utilize the GPU, which can simultaneously execute thousands of warps on the fly, we perform the graph kernel computations between many different pairs of graphs simultaneously within a single kernel launch.

One option is to assign each thread warp a unique graph pair, while the program assumes a SIMD model within each warp.
No explicit synchronization or cooperation between thread warps is needed.
It is unfavorable when low-latency computations for a few graphs are required because the work on each pair of graphs can only be parallelized among a fixed small number of threads.
Consequently, thousands of graph pairs are needed to provide enough concurrency to saturate the thousands of CUDA cores on a Volta GPU.

A second option is to further parallelize the computation within a thread block, whose size can vary between 32 to 1024 threads on CUDA GPUs.
A first and obvious benefit of this approach is that it provides us the ability to use block size to adjust the latency for computing each pair of graphs, as well as allowing a smaller number of graph pairs to saturate the entire GPU.
This block-based cooperative approach also has the potential to further improve performance by allowing warps within a block to share the octiles in shared memory.
As revealed in the Roofline analysis, larger tiles results in more data reuse, less redundant load/store operations, and higher arithmetic intensity. However, there is a limit on the size of tiles that a warp can hold without constraining occupancy, \textit{i.e.} the number of warps on the fly.
To work around this, we let all the $N$ warps in a CUDA thread block each load an octile, and then share the octiles to compute $N^2$ tile-level XMV operations.

Tile sharing requires block-level synchronization before and after octile loading.
Besides, atomic accumulations are necessary for writing to the output vector since the COO storage format obscures the effort to schedule workload among the warps in ways such that the output could be conflict-free.
However, the performance impact on CUDA GPUs should be very minimal because atomic accumulations whose outputs are not immediately used are carried out by nonblocking atomic reduction instructions. As such, the threads that commit the atomic accumulations will not get stalled.
\WISHLIST{get microbenchmark metrics to prove this, and also can try run atomics on the non-sharing solver just to measure the impact}

\WISHLIST{FIG: single-iteration block size vs performance}

\subsection{Inter-Block Load Balancing}
\label{section:load-balancing}

Thanks to the independence of the computations between different pairs of graphs, load balancing is relatively straightforward since tasks can freely relocate across thread blocks and stream processors.
Aside from transient factors such as warp scheduling, cache conflict, and atomics, the primary source of load imbalance is the variation of graph size and sparsity pattern that affect the problem size as well as the number of conjugate gradient iterations for convergence.

\WISHLIST{seems to improve at intermediate numbers of graphs.}

\section{Benchmark Dataset}
\label{section:dataset}

\subsection{Synthetic Graphs}

To test the performance of our solver, we use the Newman-Watts-Strogatz (NWS) algorithm and the Barab\'{a}si-Albert (BA) algorithm to generate synthetic graphs of small-world and scale-free characteristics, respectively.

\subsection{Real-World Dataset}

The graph kernel is further tested on real-world datasets as summarized below:
\begin{enumerate}

    \item The PDB-3k dataset is a 1324-structures subset of the Protein Data Bank database \cite{bermanProteinDataBank2000} containing proteins less than 3000 in weight and contain no DNA/RNA complexes.
    Each protein is converted into a graph with nodes representing heavy atoms.
    A spatial adjacency rule creates edges between spatially neighboring atoms such that the weights reach maximum when two atoms overlap, and smoothly decay to zero at a certain cutoff distance.
    The edges are labeled with the interatomic distance between its endpoints.

    \item DrugBank~\cite{wishartDrugBankMajorUpdate2018} is a comprehensive database containing information about drug molecules.
    The dataset contains more than $10^4$ drug molecules, 10607 of which has a corresponding linearized representation as a SMILES string, which is obtained from a depth-first traversal of the corresponding molecular graph.
    A rich body of node and edge attributes can be extracted from the SMILES strings such as hybridization state, charge, bond order, and conjugacy.

\end{enumerate}

\REM{
\begin{table}[tp!]
\centering
\footnotesize
\caption{Summary of Real-World Graph Database Used in the Benchmark.\label{table:real-graphs}}
\begin{tabularx}{\columnwidth}{@{}lXll@{}}
\toprule
\textbf{Dataset}               & \textbf{Content} & \thead[tl]{Graph\\Number} & \thead[tl]{Graph\\Size} \\ \midrule

\textbf{PDB-3k}                  & 3D protein crystal structure          & 1324                  & 30-295              \\
\textbf{DrugBank}                & Formulas of drug molecules   & 10607                 & 1-551              \\

\bottomrule
\end{tabularx}
\end{table}
}

\REM{
    PDB:
    ego-Facebook: http://snap.stanford.edu/data/ego-Facebook.html, 4000+ nodes
    gemsec-Deezer: http://snap.stanford.edu/data/gemsec-Deezer.html, 40000+ nodes
    gemsec-Facebook: http://snap.stanford.edu/data/gemsec-Facebook.html, 3000+-27000+ nodes
    arxiv Collaboration networks: GR-QC, http://snap.stanford.edu/data/ca-GrQc.html, 5242 nodes
}

\section{Performance Measurement and Analysis}
\label{section:benchmark-result}

Benchmarks are performed on the Summit supercomputer at Oak Ridge National Laboratory. The runtime and performance metrics of our GPU kernels are measured using the \texttt{nvprof} program from the CUDA Toolkit, while CPU-side time measurements are obtained using the \texttt{time.perf\_counter\_ns()} method from the Python standard library.

\subsection{Performance Improvement of Proposed Optimization Techniques}

In this section, we characterize and compare the performance gain enabled by the previously described optimization techniques on both the synthetic and real-world graph datasets. The measurements are carried out using the na\"ive kernel as a baseline and then enabling the optimization techniques one at a time in the same order as they appear in the previous sections.
For the synthetic graph datasets, we generate 160 graphs containing 96 nodes for each type with the following parameters:
\begin{itemize}[label=$\cdot$]
    \item Newman-Watts-Strogatz: $k=3$, $p=0.1$;
    \item Barab\'{a}si-Albert: $m=6$.
\end{itemize}
We also test the kernel on all graphs in the PDB and DrugBank datasets.

\begin{figure}[htp!]
  \centering
  \caption{Time to solution for the presented solver equipped with different optimization techniques (data with an asterisk are projected from ensembles of 32 random subsets of the entire dataset). Within each dataset, each bar represents a kernel that incorporates a new optimization technique while inheriting everything else from the kernel below. Label interpretation: \texttt{Dense} -- the na\"ive kernel, \texttt{Sparse} -- sparsity exploitation at the inter-tile level, \texttt{+Reorder} -- enabling PBR graph reordering, \texttt{+Adaptive} -- adaptive switching between dense and sparse tile primitives, \texttt{+Compact} -- compact storage format of each tile, \texttt{+Block}: sharing of tiles within a block, \texttt{+DynSched} -- dynamic work scheduling.
   \label{fig:incremental-speedup}
  }
  \vspace{0.5em}
  \includegraphics[width=\columnwidth]{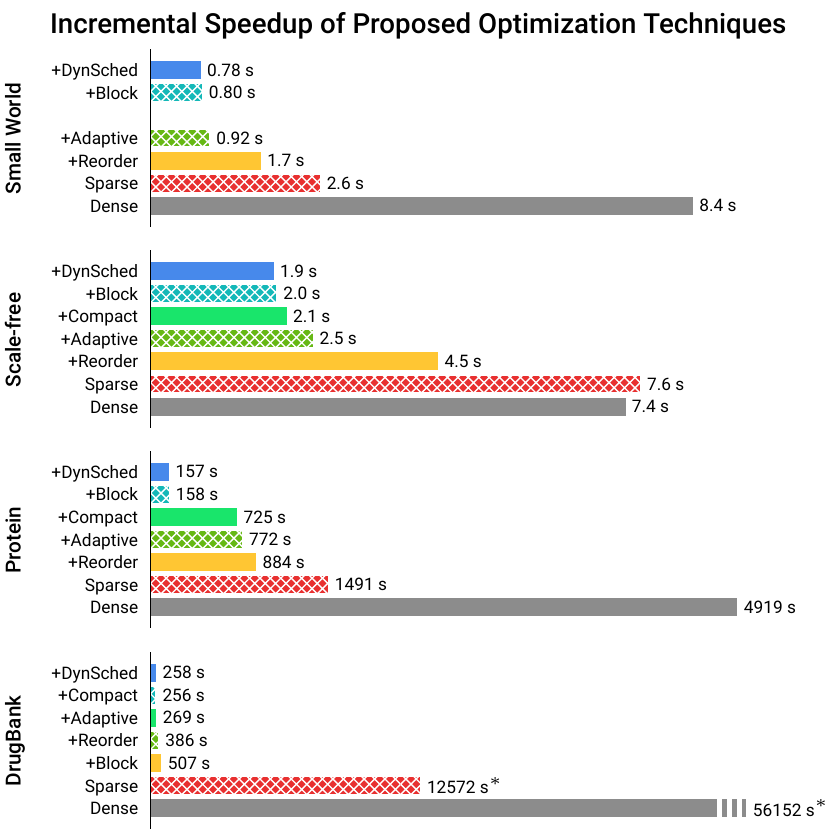}
\end{figure}

From \Cref{fig:incremental-speedup}, we can conclude that the performance improvements brought about by the techniques depend on the characteristics of the actual dataset.
Overall, the speedup is more impressive on the real-world datasets, which contain more diverse types of graphs.
It turns out inter-tile sparsity exploitation, when directly applied to the graphs in their natural order, improves the performance on all datasets except for the scale-free networks which contain poor locality.
On top of that, PBR-based reordering performs very well and increases the performance of the solver on all datasets.
The adaptive dense/sparse primitive switch and the compact tile storage format can further improve solver performance on all datasets.

Block-level tile sharing leads to significant performance improvement on DrugBank but only mild improvements on other datasets.
The reason is that only the DrugBank dataset exhibits considerable size variation with graphs containing 1 to 551 nodes. In that case, block-level tile sharing can significantly reduce the time to compute the largest pair of molecules, which otherwise takes a very long time using only a single warp.
Dynamic scheduling brings about marginal performance improvements because the GPUs are already saturated by our datasets.

\subsection{Performance Comparison with State-of-the-Art Packages}
\label{section:lateral}

We further compare the performance of our solver against two state-of-the-art packages for graph kernel computations: GraKeL ~\cite{siglidisGraKeLGraphKernel2018a} and GraphKernels~\cite{sugiyamaGraphkernelsPythonPackages2018a}.
GraKeL is a Python package compatible with scikit-learn~\cite{pedregosaScikitlearnMachineLearning2011}.
The compute-intensive part of GraKeL is implemented using Cython~\cite{behnelCythonBestBoth2011},  which compiles codes written in a Python-like syntax into binaries on the target machine.
The GraphKernels package is implemented in C++ and has a Python frontend generated with SWIG~\cite{beazleyAutomatedScientificSoftware2003}.
Both packages run only on CPUs, although GraKeL does support parallelism using multiple processes but with limited scaling efficiency.
When executing the codes on the Power9 cores of Summit, we allocate 4 physical cores to GraKeL and 1 physical core to GraphKernels.

\begin{figure}[htp!]
  \centering
  \caption{The solver presented in this paper outperforms existing Python packages by several orders of magnitude.\label{fig:vs-grakel}}
  \vspace{0.5em}
  \includegraphics[width=\columnwidth]{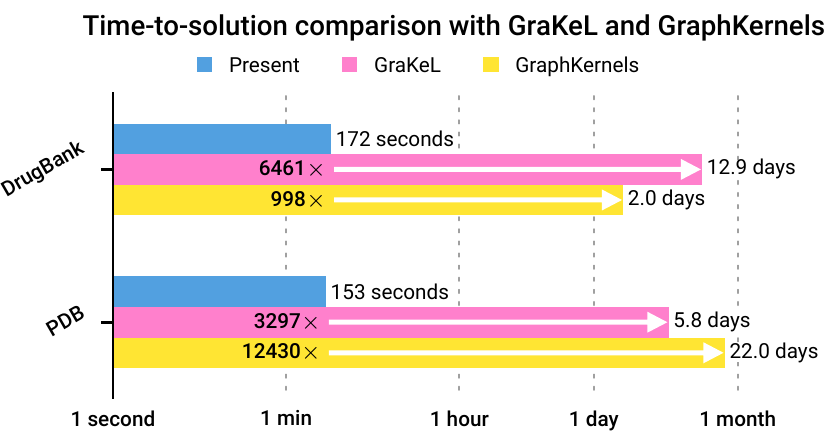}
\end{figure}

As shown by \Cref{fig:vs-grakel}, our solver significantly outperforms both GraKeL and GraphKernels by 3-4 orders of magnitude on real-world datasets. Besides performance, it is worth noting that we had to carry out the computation using a relatively large stopping probability for both GraKeL and GraphKernels to avoid convergence failures. Coincidentally, a larger stopping probability can reduce time to solution at the expense of the discriminating power of the kernel. Our presented kernel does not have a convergence issue and can compute using stopping probability values as small as 0.0005.

\section{Related Work}
\label{section:related}

The two packages GraKeL and Graph-Kernels that we have compared against in \cref{section:lateral} provide the closest functionality to our solver, but we significantly outperform them by several orders of magnitude. Moreover, only GraKeL supports graphs with both labeled vertices and labeled edges, which are crucial for building accurate machine learning models for molecular systems~\cite{tangPredictionAtomizationEnergy2019}. It is easily verifiable that the normalized Gramian matrix generated using unlabeled graphs contains only numbers all very close to unity, implying that all graphs are identical to each other under the unlabeled similarity measure. As such, we believe the present solver, which can efficiently compute the graph kernel for labeled graphs, represents not only an improvement in terms of computational speed but also an enhancement of the functionality available to the end-users.

The graph kernel is fundamentally different from network alignment algorithms~\cite{singhGlobalAlignmentMultiple2008} that can provide an estimate on the similarity of two graphs. Two issues make network alignment algorithms unsuitable as kernels between graphs. First, alignment algorithms generally are not positive definite functions, \textit{i.e.} they do not induce a norm on an associated Hilbert space. Second, alignment algorithms are potentially more expensive because it involves more work in addition to computing nodal similarities.

Last, the work of Livi \textit{et al.} \cite{liviParallelAlgorithmsTensor2012} contains an algorithmic motif concerning the parallel computation of graph tensor products for inexact graph matching. While their formulation also has a product weight matrix that is computed using a vertex kernel and an edge kernel, the product graph is not used to construct a linear system that has to be solved.

\WISHLIST{Rchemcpp \cite{klambauerRchemcppWebService2015a} }

\section{Conclusion}
\label{section:conclusion}

In this paper, we presented a series of algorithms to accelerate a marginalized graph kernel solver on GPU.
The solver is essentially an implementation of the conjugate gradient method for a generalized Laplacian system induced by the Kronecker product of a pair of graphs.
Via roofline analysis, we identified that the solver would likely be memory-bound due to the matrix-vector inner product operation in the CG method. We overcame this issue by taking advantage of the Kronecker product structure of the system.
In our approach, we do not precompute the product system, but rather stream and cache the original graph pair in tiles, and compute the product system \onthefly.
This approach significantly reduces global memory traffic, and only increases the asymptotic arithmetic operation count by a constant factor, which can be easily offset by the substantial gain in instruction throughput.
Moreover, the solver can take advantage of the sparsity in the graph by making use of a two-level storage format to trim out zero elements.
We compared the performance of the solver with existing packages and demonstrated that our implementation delivered significant speedups.

This work also exemplifies the paradigm of applying linear algebra concepts and techniques to solving graph problems.
The graph kernel problem constitutes a concrete example of the need for standardized application programming interfaces for graph tensor products in specifications such as GraphBLAS~\cite{bulucDesignGraphBLASAPI2017}, and prompt for the development of high-performance and general implementations of the interface.
Our work suggests that the semantics for the \emph{inner product} between tensor product structures may see broader applicability than that for the mere computation of the tensor product itself.

\section*{Acknowledgment}
This work was supported by the Luis W. Alvarez Postdoctoral Fellowship at Lawrence Berkeley National Laboratory. This work is also supported in part by the Applied Mathematics program of the DOE Office of Advanced Scientific Computing Research under Contract No. DE-AC02-05CH11231, and in part by the Exascale Computing Project (17-SC-20-SC), a collaborative effort of the U.S. DOE Office of Science and the NNSA.
This research used resources of the Oak Ridge Leadership Computing Facility at the Oak Ridge National Laboratory, which is supported by the Office of Science of the U.S. DOE  under Contract No. DE-AC05-00OR22725.
YHT thanks David Williams-Young, and Caitlin A Whitter for helpful discussions and suggestions.

\bibliography{bibliography}

% Generated by IEEEtran.bst, version: 1.14 (2015/08/26)
\begin{thebibliography}{10}
\providecommand{\url}[1]{#1}
\csname url@samestyle\endcsname
\providecommand{\newblock}{\relax}
\providecommand{\bibinfo}[2]{#2}
\providecommand{\BIBentrySTDinterwordspacing}{\spaceskip=0pt\relax}
\providecommand{\BIBentryALTinterwordstretchfactor}{4}
\providecommand{\BIBentryALTinterwordspacing}{\spaceskip=\fontdimen2\font plus
\BIBentryALTinterwordstretchfactor\fontdimen3\font minus
  \fontdimen4\font\relax}
\providecommand{\BIBforeignlanguage}[2]{{%
\expandafter\ifx\csname l@#1\endcsname\relax
\typeout{** WARNING: IEEEtran.bst: No hyphenation pattern has been}%
\typeout{** loaded for the language `#1'. Using the pattern for}%
\typeout{** the default language instead.}%
\else
\language=\csname l@#1\endcsname
\fi
#2}}
\providecommand{\BIBdecl}{\relax}
\BIBdecl

\bibitem{kashimaMarginalizedKernelsLabeled2003}
H.~Kashima, K.~Tsuda, and A.~Inokuchi, ``Marginalized kernels between labeled
  graphs,'' in \emph{Proceedings of the 20th International Conference on
  Machine Learning ({{ICML}}-03)}.\hskip 1em plus 0.5em minus 0.4em\relax {AAAI
  Press}, 2003, pp. 321--328, 00000.

\bibitem{tangPredictionAtomizationEnergy2019}
Y.-H. Tang and W.~A. {de Jong}, ``Prediction of atomization energy using graph
  kernel and active learning,'' \emph{The Journal of Chemical Physics}, vol.
  150, no.~4, p. 044107, Jan. 2019, autocitation-1.

\bibitem{borgwardtProteinFunctionPrediction2005a}
K.~M. Borgwardt, C.~S. Ong, S.~Sch{\"o}nauer, S.~V.~N. Vishwanathan, A.~J.
  Smola, and H.-P. Kriegel, ``\BIBforeignlanguage{en}{Protein function
  prediction via graph kernels},''
  \emph{\BIBforeignlanguage{en}{Bioinformatics}}, vol.~21, no. suppl\_1, pp.
  i47--i56, Jun. 2005.

\bibitem{tangHighThroughputSolverMarginalized2019}
Y.-H. Tang, O.~Selvitopi, D.~Popovici, and A.~Bulu{\c c}, ``A
  {{High}}-{{Throughput Solver}} for {{Marginalized Graph Kernels}} on
  {{GPU}},'' \emph{arXiv:1910.06310 [cs]}, Dec. 2019.

\bibitem{vishwanathanGraphKernels2010}
S.~V.~N. Vishwanathan, N.~N. Schraudolph, R.~Kondor, and K.~M. Borgwardt,
  ``Graph kernels,'' \emph{Journal of Machine Learning Research}, vol.~11, no.
  Apr, pp. 1201--1242, 2010, 00000.

\bibitem{vishwanathanFastComputationGraph2006}
S.~Vishwanathan, K.~M. Borgwardt, and N.~N. Schraudolph,
  ``\BIBforeignlanguage{en}{Fast {{Computation}} of {{Graph Kernels}}},'' in
  \emph{\BIBforeignlanguage{en}{{{NIPS}}}}, vol.~19, 2006, pp. 131--138.

\bibitem{jiaDissectingNVIDIAVolta2018}
Z.~Jia, M.~Maggioni, B.~Staiger, and D.~P. Scarpazza, ``Dissecting the {{NVIDIA
  Volta GPU Architecture}} via {{Microbenchmarking}},'' \emph{arXiv:1804.06826
  [cs]}, Apr. 2018.

\bibitem{williamsRooflineInsightfulVisual2009}
S.~Williams, A.~Waterman, and D.~Patterson, ``Roofline: An insightful visual
  performance model for multicore architectures,'' \emph{Communications of the
  ACM}, vol.~52, no.~4, pp. 65--76, Apr. 2009.

\bibitem{Selvitopi2017}
O.~Selvitopi, S.~Acer, and C.~Aykanat, ``A recursive hypergraph bipartitioning
  framework for reducing bandwidth and latency costs simultaneously,''
  \emph{IEEE Trans. Parallel Distrib. Syst.}, vol.~28, no.~2, pp. 345--358,
  Feb. 2017.

\bibitem{georgeComputerSolutionLarge1981a}
A.~George and J.~W.~H. Liu, \emph{\BIBforeignlanguage{eng}{Computer Solution of
  Large Sparse Positive Definite Systems}}, ser. Prentice-{{Hall}} Series in
  Computational Mathematics.\hskip 1em plus 0.5em minus 0.4em\relax {Englewood
  Cliffs, NJ}: {Prentice-Hall}, 1981.

\bibitem{pinarImprovingPerformanceSparse1999}
A.~Pinar and M.~T. Heath, ``Improving {{Performance}} of {{Sparse
  Matrix}}-{{Vector Multiplication}},'' in \emph{{{SC}} '99: {{Proceedings}} of
  the 1999 {{ACM}}/{{IEEE Conference}} on {{Supercomputing}}}, Nov. 1999, pp.
  30--30.

\bibitem{tangAcceleratingDissipativeParticle2014}
Y.-H. Tang and G.~E. Karniadakis, ``Accelerating dissipative particle dynamics
  simulations on {{GPUs}}: {{Algorithms}}, numerics and applications,''
  \emph{Computer Physics Communications}, vol. 185, no.~11, pp. 2809--2822,
  Nov. 2014, 00025.

\bibitem{sandersThinkLocallyAct2013}
P.~Sanders and C.~Schulz, ``\BIBforeignlanguage{en}{Think {{Locally}}, {{Act
  Globally}}: {{Highly Balanced Graph Partitioning}}},'' in
  \emph{\BIBforeignlanguage{en}{Experimental {{Algorithms}}}}, ser. Lecture
  {{Notes}} in {{Computer Science}}, V.~Bonifaci, C.~Demetrescu, and
  A.~{Marchetti-Spaccamela}, Eds.\hskip 1em plus 0.5em minus 0.4em\relax
  {Springer Berlin Heidelberg}, 2013, pp. 164--175.

\bibitem{Benlic2011}
U.~Benlic and J.-K. Hao, ``An effective multilevel tabu search approach for
  balanced graph partitioning,'' \emph{Comput. Oper. Res.}, vol.~38, no.~7, pp.
  1066--1075, Jul. 2011.

\bibitem{Fiduccia1982}
C.~M. Fiduccia and R.~M. Mattheyses, ``A linear-time heuristic for improving
  network partitions,'' in \emph{Proceedings of the 19th Design Automation
  Conference}, ser. {{DAC}} '82.\hskip 1em plus 0.5em minus 0.4em\relax
  {Piscataway, NJ, USA}: {IEEE Press}, 1982, pp. 175--181.

\bibitem{bermanProteinDataBank2000}
H.~M. Berman, J.~Westbrook, Z.~Feng, G.~Gilliland, T.~N. Bhat, H.~Weissig,
  I.~N. Shindyalov, and P.~E. Bourne, ``\BIBforeignlanguage{en}{The {{Protein
  Data Bank}}},'' \emph{\BIBforeignlanguage{en}{Nucleic Acids Research}},
  vol.~28, no.~1, pp. 235--242, Jan. 2000.

\bibitem{wishartDrugBankMajorUpdate2018}
D.~S. Wishart, Y.~D. Feunang, A.~C. Guo, E.~J. Lo, A.~Marcu, J.~R. Grant,
  T.~Sajed, D.~Johnson, C.~Li, Z.~Sayeeda, N.~Assempour, I.~Iynkkaran, Y.~Liu,
  A.~Maciejewski, N.~Gale, A.~Wilson, L.~Chin, R.~Cummings, D.~Le, A.~Pon,
  C.~Knox, and M.~Wilson, ``\BIBforeignlanguage{en}{{{DrugBank}} 5.0: A major
  update to the {{DrugBank}} database for 2018},''
  \emph{\BIBforeignlanguage{en}{Nucleic Acids Research}}, vol.~46, no.~D1, pp.
  D1074--D1082, Jan. 2018.

\bibitem{siglidisGraKeLGraphKernel2018a}
G.~Siglidis, G.~Nikolentzos, S.~Limnios, C.~Giatsidis, K.~Skianis, and
  M.~Vazirgianis, ``\BIBforeignlanguage{en}{{{GraKeL}}: {{A Graph Kernel
  Library}} in {{Python}}},'' Jun. 2018.

\bibitem{sugiyamaGraphkernelsPythonPackages2018a}
M.~Sugiyama, M.~E. Ghisu, F.~{Llinares-L{\'o}pez}, and K.~Borgwardt,
  ``\BIBforeignlanguage{en}{Graphkernels: {{R}} and {{Python}} packages for
  graph comparison},'' \emph{\BIBforeignlanguage{en}{Bioinformatics}}, vol.~34,
  no.~3, pp. 530--532, Feb. 2018.

\bibitem{pedregosaScikitlearnMachineLearning2011}
F.~Pedregosa, G.~Varoquaux, A.~Gramfort, V.~Michel, B.~Thirion, O.~Grisel,
  M.~Blondel, P.~Prettenhofer, R.~Weiss, V.~Dubourg, J.~Vanderplas, A.~Passos,
  D.~Cournapeau, M.~Brucher, M.~Perrot, and {\'E}.~Duchesnay, ``Scikit-learn:
  {{Machine Learning}} in {{Python}},'' \emph{Journal of Machine Learning
  Research}, vol.~12, pp. 2825--2830, Oct. 2011.

\bibitem{behnelCythonBestBoth2011}
S.~Behnel, R.~Bradshaw, C.~Citro, L.~Dalcin, D.~S. Seljebotn, and K.~Smith,
  ``Cython: {{The Best}} of {{Both Worlds}},'' \emph{Computing in Science and
  Engg.}, vol.~13, no.~2, pp. 31--39, Mar. 2011.

\bibitem{beazleyAutomatedScientificSoftware2003}
D.~M. Beazley, ``Automated {{Scientific Software Scripting}} with {{SWIG}},''
  \emph{Future Gener. Comput. Syst.}, vol.~19, no.~5, pp. 599--609, Jul. 2003.

\bibitem{singhGlobalAlignmentMultiple2008}
R.~Singh, J.~Xu, and B.~Berger, ``\BIBforeignlanguage{en}{Global alignment of
  multiple protein interaction networks with application to functional
  orthology detection},'' \emph{\BIBforeignlanguage{en}{Proceedings of the
  National Academy of Sciences}}, vol. 105, no.~35, pp. 12\,763--12\,768, Sep.
  2008.

\bibitem{liviParallelAlgorithmsTensor2012}
L.~Livi and A.~Rizzi, ``Parallel algorithms for tensor product-based inexact
  graph matching,'' in \emph{The 2012 {{International Joint Conference}} on
  {{Neural Networks}} ({{IJCNN}})}, Jun. 2012, pp. 1--8.

\bibitem{bulucDesignGraphBLASAPI2017}
A.~Bulu{\c c}, T.~Mattson, S.~McMillan, J.~Moreira, and C.~Yang, ``Design of
  the {{GraphBLAS API}} for {{C}},'' in \emph{2017 {{IEEE International
  Parallel}} and {{Distributed Processing Symposium Workshops}} ({{IPDPSW}})},
  May 2017, pp. 643--652.

\bibitem{wendlandScatteredDataApproximation2004}
{Wendland}, \emph{Scattered {{Data Approximation}}}.\hskip 1em plus 0.5em minus
  0.4em\relax {Cambridge university press}, 2004, vol.~17.

\bibitem{tangAtomisticFingerprintAlgorithm2018}
Y.-H. Tang, D.~Zhang, and G.~E. Karniadakis, ``An atomistic fingerprint
  algorithm for learning ab initio molecular force fields,'' \emph{The Journal
  of Chemical Physics}, vol. 148, no.~3, p. 034101, Jan. 2018, 00000.

\end{thebibliography}
\bibliographystyle{IEEEtran}

\channelifnotthen{ipdps}{

\clearpage

\onecolumn
\appendix

\titlespacing*{\section}{0pt}{1.0ex}{1.0ex}
\titlespacing*{\subsection}{0pt}{.75ex}{0.75ex}
\titleformat{\paragraph}[block]{\sffamily\bfseries\filcenter}{}{0.5em}{}

\subsection{Derivation of the Linear Algebra Form of the Marginalized Graph Kernel}\label{section:derivation-of-the-linear-algebra-form-of-the-marginalized-graph-kernel}

A formula for evaluating the marginalized graph kernel, as directly implied by the random walk picture, reads:
\begin{align}
    K(G,G') = \sum_{\ell=1}^\infty
              \sum_{\mathbf{h}}
              \sum_{\mathbf{h}'}
              \Bigg[
                &\ p_s(h_1)\ p'_s(h'_1)\ \kappa_\mathrm{v}(v_{h_1}, v'_{h'_1})\  
                p_q(h_\ell)\ p'_q(h'_\ell)\ \left( \prod_{i=2}^\ell p_t(h_i|h_{i-1}) \right)\ 
                \left( \prod_{j=2}^\ell p'_t(h'_j|h'_{j-1}) \right) \nonumber\\
                & \left( \prod_{k=2}^\ell \kappa_\mathrm{v}(v_{h_k}, v'_{h'_k}) \kappa_\mathrm{e}( e_{h_{k-1} h_k}, e'_{h'_{k-1} h'_k} )\right) \
              \Bigg]. \label{eq:mlgk-summation-form}
\end{align}

However, an equivalent formulation \cite{tangPredictionAtomizationEnergy2019}, which transforms the task into solving a generalized Laplacian equation on the tensor product graph, permits more efficient numerical computation. To obtain this linear algebra formulation, we restate \Cref{eq:mlgk-summation-form} under the spirit of dynamic programming following \cite{kashimaMarginalizedKernelsLabeled2003}:

\begin{equation}
  K(G,G') = \sum_{h_1 \in V, h'_1 \in V'} p_s(h_1)\ p'_s(h'_1)\ \kappa_\mathrm{v}(h_1, h'_1)\ R_\infty(h_1, h'_1), \label{eq:mlgk-as-sum-over-R}
\end{equation}
where $R_\infty$ is the solution to the linear system:
\begin{equation}
R_\infty(h_1, h'_1) = p_q(h_1)\ p'_q(h'_1) + \sum_{i \in V,j \in V'} t(i,j,h_1,h'_1)\ R_\infty(i,j), \label{eq:mlgk-R-infty}
\end{equation}
with
\begin{equation}
t(i,j,h_1,h'_1) \coloneqq p_t(i|h_1)\ p'_t(j|h'_1)\ \kappa_\mathrm{v}(v_i,v_j)\ \kappa_\mathrm{e}( e_{i\,h_1}, e_{j\,h'_1} ).\label{eq:mlgk-t-definition}
\end{equation}

\Cref{eq:mlgk-as-sum-over-R,eq:mlgk-R-infty,eq:mlgk-t-definition} exhibit a Kronecker product structure, which can be readily recognized in matrix form:

\begin{equation}
  K(G,G') = \left( \mathbf{p} \otimes \vphantom{\overset{\kappa_\mathrm{v}}{\smallotimes}} \mathbf{p}' \right)^\mathsf{T} \cdot \mathbf{diag}\left( \mathbf{v} \overset{\kappa_\mathrm{v}}{\smallotimes} \mathbf{v}' \right) \cdot \mathbf{r}_\infty, \label{eq:mlgk-linear-form}
\end{equation}
with $\mathbf{r}_\infty$ being the solution to the linear system
\begin{equation}
  \mathbf{r}_\infty = \mathbf{q} \otimes \mathbf{q}' + \left[ \left( \mathbf{P} \otimes \mathbf{P}' \vphantom{\overset{\kappa_\mathrm{v}}{\smallotimes}} \right) \odot \left( \mathbf{E} \overset{\kappa_\mathrm{e}}{\smallotimes} \mathbf{E}' \right) \right] \cdot \mathbf{diag}\left( \mathbf{v} \overset{\kappa_\mathrm{v}}{\smallotimes} \mathbf{v}' \right) \cdot \mathbf{r}_\infty, \label{eq:r-infinity-linear-system}
\end{equation}
where
\begin{itemize}[label=,leftmargin=8em]
  \item[$\mathbf{v}\phantom{'}$] is the vertex label vector of $G$ with $\mathbf{v}_i = v_i$;
  \item[$\mathbf{p}\phantom{'}$] is the starting probability vector of $G$ with $\mathbf{p}_i = p_s(v_i)$;
  \item[$\mathbf{q}\phantom{'}$] is the stopping probability vector of $G$ with $\mathbf{q}_i = p_q(v_i)$;
  \item[$\mathbf{P}\phantom{'}$] is the transition probability matrix of $G$ defined as $\mathbf{D}^{-1} \mathbf{A}$;
  \item[$\mathbf{E}\phantom{'}$] is the edge label matrix of $G$ with $\mathbf{E}_{ij} = e_{ij}$;
  \item[$\mathbf{v}'$, $\mathbf{p}'$, $\mathbf{q}'$, $\mathbf{P}'$, $\mathbf{E}'$] are the corresponding vectors and matrices for $G'$;
  \item[$\overset{\kappa_\mathrm{v}}{\smallotimes}\phantom{'}$] is the generalized Kronecker product between $\mathbf{v}$ and $\mathbf{v}'$ with respect to $\kappa_\mathrm{v}$;
  \item[$\overset{\kappa_\mathrm{e}}{\smallotimes}\phantom{'}$] is the generalized Kronecker product between $\mathbf{E}$ and $\mathbf{E}'$ with respect to $\kappa_\mathrm{e}$.
\end{itemize}

\vspace{0.5em}

For clarity of discussion, we denote
\begin{align*}
    \mathbf{V}_\times & \coloneqq \mathbf{diag}\left( \mathbf{v} \overset{\kappa_\mathrm{v}}{\smallotimes} \mathbf{v}' \right), \\
    \mathbf{D}_\times & \coloneqq \mathbf{diag}(\mathbf{d}) \otimes \mathbf{diag}(\mathbf{d}'), \\
    \mathbf{A}_\times & \coloneqq \mathbf{A} \otimes \mathbf{A}', \\
    \mathbf{P}_\times & \coloneqq \mathbf{P} \otimes \mathbf{P}' = \mathbf{D}_\times^{-1} \mathbf{A}_\times, \\
    \mathbf{E}_\times & \coloneqq \mathbf{E} \overset{\kappa_\mathrm{e}}{\smallotimes} \mathbf{E}', \\
    \mathbf{p}_\times & \coloneqq \mathbf{p} \otimes \mathbf{p}', \\
    \mathbf{q}_\times & \coloneqq \mathbf{q} \otimes \mathbf{q}'.
\end{align*}

To solve \cref{eq:r-infinity-linear-system}, first observe that only the product $\mathbf{V}_\times \mathbf{r}_\infty$ as a whole is needed to compute $K(G,G')$. We can thus rearrange \cref{eq:r-infinity-linear-system} to form a symmetric linear system.
\begin{align}
\mathbf{r}_\infty - \left( \mathbf{P}_\times \odot \mathbf{E}_\times \right)  \mathbf{V}_\times \mathbf{r}_\infty &= \mathbf{q}_\times,\\
\left( \mathbf{V}_\times^{-1} - \mathbf{P}_\times \odot \mathbf{E}_\times \right)  \mathbf{V}_\times \mathbf{r}_\infty &= \mathbf{q}_\times,\\
\mathbf{V}_\times \mathbf{r}_\infty &= \left( \mathbf{V}_\times^{-1} - \mathbf{P}_\times \odot \mathbf{E}_\times \right)^{-1} \mathbf{q}_\times\\
&= \left[ \mathbf{V}_\times^{-1} - \left( \mathbf{D}_\times^{-1}\mathbf{A}_\times \right)  \odot \mathbf{E}_\times \right]^{-1} \mathbf{q}_\times\\
&= \left( \mathbf{D}_\times \mathbf{V}_\times^{-1} - \mathbf{A}_\times \odot \mathbf{E}_\times \right)^{-1} \mathbf{D}_\times \mathbf{q}_\times. \label{eq:VR-symmetric-linear-form}
\end{align}

The linear system $\mathbf{D}_\times \mathbf{V}_\times^{-1} - \mathbf{A}_\times \odot \mathbf{E}_\times$ in \cref{eq:VR-symmetric-linear-form} is symmetric and positive-definite, as long as $q>0$, $\kappa_\mathrm{v}(\cdot, \cdot) <= 1$, and $\kappa_\mathrm{e}(\cdot, \cdot)<=1$.
Thus, we have reached the full expression for the marginalized graph kernel in matrix form, the solution of which is the central focus of this paper:
\begin{equation}
K(G,G') = \mathbf{p}_\times^\mathsf{T} \left( \mathbf{D}_\times \mathbf{V}_\times^{-1} - \mathbf{A}_\times \odot \mathbf{E}_\times \right)^{-1} \mathbf{D}_\times \mathbf{q}_\times.
\end{equation}

\subsection{Example of edge label and kernel}\label{section:edge-label-and-kernel-example}

Some examples of edge kernels used in practice are: 1) a square exponential kernel $\kappa^\mathrm{SE}(e_1, e_2) \doteq \exp[ -\alpha ( e_1 - e_2 )^2 ]$ consumes two floats and carries out 3 multiplication and 1 exponentiation; 2) a degree $n$ compact polynomial radial basis kernel, \textit{e.g.} in the form $\kappa^\mathrm{P}(e_1, e_2) \doteq \sum_i \alpha_i (e_1 - e_2)^i $ \cite{wendlandScatteredDataApproximation2004,tangAtomisticFingerprintAlgorithm2018} consumes two floats and performs $n$ chained FMA instructions; 3) a Kronecker product kernel $\kappa^\mathrm{kron}(e_1, e_2) \doteq \prod_i \kappa_i( e_1^i, e_2^i ) $ consumes $2n$ inputs and carry out a linearly proportional number of operations; 4) an R-convolutional kernel $\kappa^\mathrm{R}(e_1, e_2) = \sum_i \sum_j \kappa( e_1^i, e_2^j )$ consumes $2n$ inputs and carry out a quadratically proportional number of arithmetics.

\newcounter{linenoNAIVE}
\newcounter{linenoST}
\newcounter{linenoRB}
\newcounter{linenoSTRB}
\newcommand{\costtitle}{
    \thead[tr]{Line} & \thead[tc]{Algorithm} & \thead[tc]{Category} & \thead[tc]{Loop count} & \thead[tc]{Unit cost} & \thead[tc]{Total cost}
}
\newcommand{\srcline}[5]{}
\newcommand{\parfor}[1]{\textbf{parfor} #1 \textbf{do}}
\newcommand{\for}[1]{\textbf{for} #1 \textbf{do}}
\newcommand{\LDG}{LD.G}
\newcommand{\STG}{ST.G}
\newcommand{\LDS}{LD.S}
\newcommand{\STS}{ST.S}
\newcommand{\OPS}{OPS}
\newcommand{\ditto}{
    \tikz{
        \draw [line width=0.12ex] (-0.2ex,0) -- +(0,0.8ex)
            (0.2ex,0) -- +(0,0.8ex);
        \draw [line width=0.08ex] (-0.6ex,0.4ex) -- +(-0.4em,0)
            (0.6ex,0.4ex) -- +(0.4em,0);
    }
}

\newenvironment{costlisting}
{
    \begin{center}
        \vspace{-0.5em}
        
        \setlength{\tabcolsep}{0.01\textwidth}
        \setlength{\tabulinesep}{2.5pt}
        \begin{longtabu}to\textwidth{
            >{\raggedleft\arraybackslash}m{0.03\textwidth}
            >{\raggedright\arraybackslash}m{0.42\textwidth}
            >{\centering\arraybackslash}m{0.07\textwidth}
            >{\centering\arraybackslash}m{0.20\textwidth}
            >{\centering\arraybackslash}m{0.09\textwidth}
            >{\centering\arraybackslash}m{0.09\textwidth}}
            \toprule
            \costtitle \\
            \midrule
        \endfirsthead
            \multicolumn{6}{c}
            {\tablename\ \thetable\ -- \textit{Continued from previous page}} \\
            \midrule
            \costtitle \\
            \midrule
        \endhead
            \bottomrule
            \multicolumn{6}{r}{
              \textit{Continued on next page}
            }\\
        \endfoot
            \bottomrule
        \endlastfoot
}
{
        \end{longtabu}
    \end{center}
}

\subsection{Pseudocode, I/O and Operation Counts of \onthefly XMV primitives}
\label{section:primitives}

\paragraph{Na\"ive}

\renewcommand{\srcline}[5]{
    \refstepcounter{linenoNAIVE}{\sffamily\color[rgb]{0.15, 0.35, 0.9}\footnotesize\thelinenoNAIVE} & #1 & \ifblank{#3}{}{\badge{#3}} & #2 & #4 & #5
}
\begin{costlisting}
    \srcline{\hspace{0em}\parfor{$i \in [0 \ldots n m]$}}{}{}{}{} \\
    \srcline{\hspace{1em}   $\mathbf{a}_{i} \gets 0$}{}{}{}{} \\
    \srcline{\hspace{1em}   \for{$J \in [0, \warpSize, \ldots n m]$}}{}{}{}{} \\
    \label{line:FULL-load-p}
    \srcline{\hspace{2em}       \Call{GlobalLoad}{$\mathbf{p}_{J + \lane}$}}{$nm \cdot \sfrac{nm}{32}$}{\LDG}{$F$}{$\sfrac{n^2 m^2 F}{32}$} \\
    \srcline{\hspace{2em}       \for{$j \in J + [0, 32)$}}{}{}{}{} \\
    \label{line:FULL-load-L}
    \srcline{\hspace{3em}           \Call{GlobalLoad}{$\mathbf{L}_{ij}$}}{$nm \cdot nm$}{\LDG}{$F$}{$n^2 m^2 F$} \\
    \srcline{\hspace{3em}           \Call{Shuffle}{$\mathbf{p}_{j}$} from lane $j$}{}{}{}{} \\
    \label{line:FULL-compute}
    \srcline{\hspace{3em}           $\mathbf{a}_{i} \gets \mathbf{a}_{i} + \mathbf{L}_{ij} \times \mathbf{p}_{j}$}{$nm \cdot nm$}{\OPS}{2}{$2 n^2 m^2$} \\
    \label{line:FULL-store-lhs}
    \srcline{\hspace{1em}   \Call{GlobalStore}{$\mathbf{a}_{i}$}}{$nm$}{\STG}{$F$}{$n m F$} \\
\end{costlisting}

\paragraph{Shared Tiling}
\label{section:appendix-shared-tiling-primitive}

\renewcommand{\srcline}[5]{
    \refstepcounter{linenoST}{\sffamily\color[rgb]{0.15, 0.35, 0.9}\footnotesize\thelinenoST} & #1 & \ifblank{#3}{}{\badge{#3}} & #2 & #4 & #5
}
\begin{costlisting}
    \srcline{\hspace{0em}\for{$I \in [0, t, 2t \ldots n), I' \in [0, t, 2t, \ldots m)$}}{}{}{}{} \\
    \srcline{\hspace{1em}   \parfor{$i \in I + [0, t), i' \in I' + [0, t)$}}{}{}{}{} \\
    \srcline{\hspace{2em}       $\mathbf{a}_{ii'} \gets 0$}{}{}{}{} \\
    \srcline{\hspace{1em}   \for{$J \in [0, r, 2r \ldots n)$}}{}{}{}{} \\
    \label{line:STX-load-A1}
    \srcline{\hspace{2em}       \Call{GlobalLoad}{$\mathbf{A}_{I + [0, t), J + [0, r)}$}}{$\frac{n}{t}\cdot\frac{m}{t}\cdot\frac{n}{r}$}{\LDG}{$r t F$}{$\sfrac{n^2 m F}{t}$} \\
    \label{line:STX-store-A1}
    \srcline{\hspace{2em}       \Call{SharedStore}{$\mathbf{A}_{I + [0, t), J + [0, r)}$}}{\ditto}{\STS}{$r t F$}{$\sfrac{n^2 m F}{t}$} \\
    \label{line:STX-load-E1}
    \srcline{\hspace{2em}       \Call{GlobalLoad}{$\mathbf{E}_{I + [0, t), J + [0, r)}$}}{\ditto}{\LDG}{$r t E$}{$\sfrac{n^2 m E}{t}$} \\
    \label{line:STX-store-E1}
    \srcline{\hspace{2em}       \Call{SharedStore}{$\mathbf{E}_{I + [0, t), J + [0, r)}$}}{\ditto}{\STS}{$r t E$}{$\sfrac{n^2 m E}{t}$} \\
    \srcline{\hspace{2em}       \for{$J' \in [0, r, 2r \ldots m)$}}{}{}{}{} \\
    \label{line:STX-load-A2}
    \srcline{\hspace{3em}           \Call{GlobalLoad}{$\mathbf{A}'_{I' + [0, t), J' + [0, r)}$}}{$\frac{n}{t}\cdot\frac{m}{t}\cdot\frac{n}{r}\cdot\frac{m}{r}$}{\LDG}{$r t F$}{$\sfrac{n^2 m^2 F}{r t}$} \\
    \label{line:STX-store-A2}
    \srcline{\hspace{3em}           \Call{SharedStore}{$\mathbf{A}'_{I' + [0, t), J' + [0, r)}$}}{\ditto}{\STS}{$r t F$}{$\sfrac{n^2 m^2 F}{r t}$} \\
    \label{line:STX-load-E2}
    \srcline{\hspace{3em}           \Call{GlobalLoad}{$\mathbf{E}'_{I' + [0, t), J' + [0, r)}$}}{\ditto}{\LDG}{$r t E$}{$\sfrac{n^2 m^2 E}{r t}$} \\
    \label{line:STX-store-E2}
    \srcline{\hspace{3em}           \Call{SharedStore}{$\mathbf{E}'_{I' + [0, t), J' + [0, r)}$}}{\ditto}{\STS}{$r t E$}{$\sfrac{n^2 m^2 E}{r t}$} \\
    \label{line:STX-load-p}
    \srcline{\hspace{3em}           \Call{GlobalLoad}{$\mathbf{p}_{J+[0, r), J'+[0, r)}$}}{\ditto}{\LDG}{$r^2 F$}{$\sfrac{n^2 m^2 F}{t^2}$} \\
    \label{line:STX-store-p}
    \srcline{\hspace{3em}           \Call{SharedStore}{$\mathbf{p}_{J+[0, r), J'+[0, r)}$}}{\ditto}{\STS}{$r^2 F$}{$\sfrac{n^2 m^2 F}{t^2}$} \\
    \srcline{\hspace{3em}           \parfor{$i \in I + [0, t), i' \in I' + [0, t)$}}{}{}{}{} \\
    \srcline{\hspace{4em}               \for{$j \in J + [0, r)$}}{}{}{}{} \\
    \label{line:STX-load-a1e1}
    \srcline{\hspace{5em}                   \Call{SharedLoad}{$\mathbf{A}_{ij}, \mathbf{E}_{ij}$}}{$\frac{n}{t}\cdot\frac{m}{t}\cdot\frac{n}{r}\cdot\frac{m}{r}\cdot t \cdot t \cdot r$}{\LDS}{$E+F$}{$\sfrac{n^2 m^2 (E + F)}{r}$} \\
    \srcline{\hspace{5em}                   \for{$j' \in J' + [0, r)$}}{}{}{}{} \\
    \label{line:STX-load-a2}
    \srcline{\hspace{6em}                       \Call{SharedLoad}{$\mathbf{A}_{i'j'}$}}{$\frac{n}{t}\cdot\frac{m}{t}\cdot\frac{n}{r}\cdot\frac{m}{r}\cdot t \cdot t \cdot r \cdot r$}{\LDS}{$F$}{$n^2 m^2 F$} \\
    \label{line:STX-load-e2}
    \srcline{\hspace{6em}                       \Call{SharedLoad}{$\mathbf{E}_{i'j'}$}}{\ditto}{\LDS}{$E$}{$n^2 m^2 E$} \\
    \label{line:STX-load-rhs}
    \srcline{\hspace{6em}                       \Call{SharedLoad}{$\mathbf{p}_{jj'}$}}{\ditto}{\LDS}{$F$}{$n^2 m^2 F$} \\
    \srcline{\hspace{6em}                       $\mathbf{a}_{ii'} \gets \mathbf{a}_{ii'} + \mathbf{p}_{jj'} \cdot \mathbf{A}_{ij} \cdot \mathbf{A}'_{i'j'}$}{}{}{}{} \\
    \label{line:STX-compute}
    \srcline{\hspace{6em}                       $\phantom{\mathbf{a}_{ii'} }\cdot \kappa_\mathrm{e}(\mathbf{E}_{ij}, \mathbf{E}'_{i'j'})$}{\ditto}{\OPS}{$X$}{$n^2 m^2 X$} \\
    \label{line:STX-store-lhs}
    \srcline{\hspace{1em}   \Call{GlobalStore}{$\mathbf{a}_{I + [0, t), I' + [0, t)}$}}{$\frac{n}{t} \cdot \frac{m}{t}$}{\STG}{$t^2 F$}{$n m F$} \\
\end{costlisting}

\paragraph{Register Blocking}
\label{section:appendix-register-blocking-primitive}

\renewcommand{\srcline}[5]{
    \refstepcounter{linenoRB}{\sffamily\color[rgb]{0.15, 0.35, 0.9}\footnotesize\thelinenoRB} & #1 & \ifblank{#3}{}{\badge{#3}} & #2 & #4 & #5
}
\begin{costlisting}
    \srcline{\hspace{0em}\for{$I \in [0, t, 2t \ldots n), I' \in [0, t, 2t, \ldots m)$}}{}{}{}{} \\
    \srcline{\hspace{1em}   $\mathbf{a}_{I + [0, t), I' + [0, t)} \gets 0$}{}{}{}{} \\
    \srcline{\hspace{1em}   \for{$J \in [0, r, 2r \ldots n)$}}{}{}{}{} \\
    \label{line:RBX-load-A1}
    \srcline{\hspace{2em}       \Call{GlobalLoad}{$\mathbf{A}_{I + [0, t), J + [0, r)}$}}{$\frac{n}{t}\cdot\frac{m}{t}\cdot\frac{n}{r}$}{\LDG}{$r t F$}{$\sfrac{n^2 m F}{t}$} \\
    \label{line:RBX-load-E1}
    \srcline{\hspace{2em}       \Call{GlobalLoad}{$\mathbf{E}_{I + [0, t), J + [0, r)}$}}{\ditto}{\LDG}{$r t E$}{$\sfrac{n^2 m E}{t}$} \\
    \srcline{\hspace{2em}       \for{$J' \in [0, r, 2r, \ldots m)$}}{}{}{}{} \\
    \label{line:RBX-load-A2}
    \srcline{\hspace{3em}           \Call{GlobalLoad}{$\mathbf{A}'_{I' + [0, t), J' + [0, r)}$}}{$\frac{n}{t}\cdot\frac{m}{t}\cdot\frac{n}{r}\cdot\frac{m}{r}$}{\LDG}{$r t F$}{$\sfrac{n^2 m^2 F}{rt}$} \\
    \label{line:RBX-load-E2}
    \srcline{\hspace{3em}           \Call{GlobalLoad}{$\mathbf{E}'_{I' + [0, t), J' + [0, r)}$}}{\ditto}{\LDG}{$r t E$}{$\sfrac{n^2 m^2 E}{rt}$} \\
    \label{line:RBX-load-p}
    \srcline{\hspace{3em}           \Call{GlobalLoad}{$\mathbf{p}_{J+[0,r), J'+[0,r]}$}}{\ditto}{\LDG}{$r^2 F$}{$\sfrac{n^2 m^2 F}{t^2}$} \\
    \label{line:RBX-store-p}
    \srcline{\hspace{3em}           \Call{SharedStore}{$\mathbf{p}_{J+[0,r), J'+[0,r]}$} }{\ditto}{\STS}{$r^2 F$}{$\sfrac{n^2 m^2 F}{t^2}$} \\
    \srcline{\hspace{3em}           \parfor{$i \in I + [0, t), i' \in I' + [0, t)$}}{}{}{}{} \\
    \srcline{\hspace{4em}               \for{$j \in J + [0, r), j' \in J' + [0, r)$}}{}{}{}{} \\
    \label{line:RBX-load-rhs}
    \srcline{\hspace{5em}                   \Call{SharedLoad}{$\mathbf{p}_{jj'}$}}{$n^2 m^2$}{\LDS}{$F$}{$n^2 m^2 F$} \\
    \srcline{\hspace{6em}                       $\mathbf{a}_{ii'} \gets \mathbf{a}_{ii'} + \mathbf{p}_{jj'} \cdot \mathbf{A}_{ij} \cdot \mathbf{A}'_{i'j'}$}{}{}{}{} \\
    \label{line:RBX-compute}
    \srcline{\hspace{6em}                       $\phantom{\mathbf{a}_{ii'} }\cdot \kappa_\mathrm{e}(\mathbf{E}_{ij}, \mathbf{E}'_{i'j'})$}{\ditto}{\OPS}{$X$}{$n^2 m^2 X$} \\
    \label{line:RBX-store-lhs}
    \srcline{\hspace{1em}   \Call{GlobalStore}{$\mathbf{a}_{I + [0, t), I' + [0, t')}$} }{$\frac{n}{t}\cdot\frac{m}{t}$}{\STG}{$t^2 F$}{$n m F$} \\
\end{costlisting}

\paragraph{Tiling-Blocking}
\label{section:appendix-tiling-blocking-primitive}

\renewcommand{\srcline}[5]{
    \refstepcounter{linenoSTRB}{\sffamily\color[rgb]{0.15, 0.35, 0.9}\footnotesize\thelinenoSTRB} & #1 & \ifblank{#3}{}{\badge{#3}} & #2 & #4 & #5
}
\begin{costlisting}
    \srcline{\hspace{0em}\for{$I \in [0, t, 2t \ldots n), I' \in [0, t, 2t, \ldots m)$}}{}{}{}{} \\
    \srcline{\hspace{1em}    \parfor{$i \in I + [0, t), i' \in I' + [0, t)$}}{}{}{}{} \\
    \srcline{\hspace{2em}        $\mathbf{a}_{ii'} \gets 0$}{}{}{}{} \\
    \srcline{\hspace{1em}    \for{$J \in [0, t, 2t \ldots n)$}}{}{}{}{} \\
    \label{line:STRBX-load-A1}
    \srcline{\hspace{2em}        \Call{GlobalLoad}{$\mathbf{A}_{I + [0, t), J + [0, t)}$}}{$\frac{n}{t}\cdot\frac{m}{t}\cdot\frac{n}{t}$}{\LDG}{$t^2 F$}{$\sfrac{n^2 m F}{t}$} \\
    \label{line:STRBX-store-A1}
    \srcline{\hspace{2em}        \Call{SharedStore}{$\mathbf{A}_{I + [0, t), J + [0, t)}$}}{\ditto}{\STS}{$t^2 F$}{$\sfrac{n^2 m F}{t}$} \\
    \label{line:STRBX-load-E1}
    \srcline{\hspace{2em}        \Call{GlobalLoad}{$\mathbf{E}_{I + [0, t), J + [0, t)}$}}{\ditto}{\LDG}{$t^2 E$}{$\sfrac{n^2 m E}{t}$} \\
    \label{line:STRBX-store-E1}
    \srcline{\hspace{2em}        \Call{SharedStore}{$\mathbf{E}_{I + [0, t), J + [0, t)}$}}{\ditto}{\STS}{$t^2 E$}{$\sfrac{n^2 m E}{t}$} \\
    \srcline{\hspace{2em}        \for{$J' \in [0, t, 2t \ldots m)$}}{}{}{}{} \\
    \label{line:STRBX-load-A2}
    \srcline{\hspace{3em}            \Call{GlobalLoad}{$\mathbf{A}'_{I' + [0, t), J' + [0, t)}$}}{$\frac{n}{t}\cdot\frac{m}{t}\cdot\frac{n}{t}\cdot\frac{m}{t}$}{\LDG}{$t^2 F$}{$\sfrac{n^2 m^2 F}{t^2}$} \\
    \label{line:STRBX-store-A2}
    \srcline{\hspace{3em}            \Call{SharedStore}{$\mathbf{A}'_{I' + [0, t), J' + [0, t)}$}}{\ditto}{\STS}{$t^2 F$}{$\sfrac{n^2 m^2 F}{t^2}$} \\
    \label{line:STRBX-load-E2}
    \srcline{\hspace{3em}            \Call{GlobalLoad}{$\mathbf{E}'_{I' + [0, t), J' + [0, t)}$}}{\ditto}{\LDG}{$t^2 E$}{$\sfrac{n^2 m^2 E}{t^2}$} \\
    \label{line:STRBX-store-E2}
    \srcline{\hspace{3em}            \Call{SharedStore}{$\mathbf{E}'_{I' + [0, t), J' + [0, t)}$}}{\ditto}{\STS}{$t^2 E$}{$\sfrac{n^2 m^2 E}{t^2}$} \\
    \label{line:STRBX-load-p}
    \srcline{\hspace{3em}            \Call{GlobalLoad}{$\mathbf{p}_{J+[0, t), J'+[0, t)}$}}{\ditto}{\LDG}{$t^2 F$}{$\sfrac{n^2 m^2 F}{t^2}$} \\
    \srcline{\hspace{3em}            \parfor{$i \in I + [0, t), i' \in I' + [0, t)$}}{}{}{}{} \\
    \srcline{\hspace{4em}                \for{$h \in [J, J+r, \ldots J+t)$}}{}{}{}{} \\
    \label{line:STRBX-load-a1}
    \srcline{\hspace{5em}                    \Call{SharedLoad}{$\mathbf{A}_{i,h+[0,r)}$}}{$\frac{n}{t}\cdot\frac{m}{t}\cdot\frac{n}{t}\cdot\frac{m}{t}\cdot t \cdot t \cdot \frac{t}{r}$}{\LDS}{$r F$}{$\sfrac{n^2 m^2 F}{t}$} \\
    \label{line:STRBX-load-e1}
    \srcline{\hspace{5em}                    \Call{SharedLoad}{$\mathbf{E}_{i,h+[0,r)}$}}{\ditto}{\LDS}{$r E$}{$\sfrac{n^2 m^2 E}{t}$} \\
    \srcline{\hspace{5em}                    \for{$h' \in [J', J'+r, \ldots J'+t)$}}{}{}{}{} \\
    \label{line:STRBX-load-a2}
    \srcline{\hspace{6em}                        \Call{SharedLoad}{$\mathbf{A}'_{i',h'+[0,r)}$}}{$\frac{n}{t}\cdot\frac{m}{t}\cdot\frac{n}{t}\cdot\frac{m}{t}\cdot t \cdot t \cdot \frac{t}{r} \cdot \frac{t}{r}$}{\LDS}{$r F$}{$\sfrac{n^2 m^2 F}{r}$} \\
    \label{line:STRBX-load-e2}
    \srcline{\hspace{6em}                        \Call{SharedLoad}{$\mathbf{E}'_{i',h'+[0,r)}$}}{\ditto}{\LDS}{$r E$}{$\sfrac{n^2 m^2 E}{r}$} \\
    \srcline{\hspace{6em}                        \for{$j \in h + [0, r)$}}{}{}{}{} \\
    \srcline{\hspace{7em}                            \for{$j' \in h' + [0, r)$}}{}{}{}{} \\
    \srcline{\hspace{8em}                                $\mathbf{a}_{ii'} \gets \mathbf{a}_{ii'} + \mathbf{p}_{jj'} \cdot \mathbf{A}_{ij} \cdot \mathbf{A}'_{i'j'}$}{}{}{}{} \\
    \label{line:STRBX-compute}
    \srcline{\hspace{8em}                                $\phantom{\mathbf{a}_{ii'} }\cdot \kappa_\mathrm{e}(\mathbf{E}_{ij}, \mathbf{E}'_{i'j'})$}{$n^2 m^2$}{\OPS}{$X$}{$n^2 m^2 X$} \\
    \label{line:STRBX-store-lhs}
    \srcline{\hspace{1em}   \Call{GlobalStore}{$\mathbf{a}_{I + [0, t), I' + [0, t)}$}}{$\dfrac{n}{t}\cdot\dfrac{m}{t}$}{\STG}{$t^2 F$}{$n m F$} \\
\end{costlisting}
}

\end{document}